\documentclass[onecolumn]{aastex631}
\usepackage{graphicx}
\usepackage{ulem}
\newcommand{\sw}[1]{\texttt{#1}}

\shorttitle{DSA-110 FRB Sample}
\shortauthors{et al.}
\graphicspath{{./}{figures/}}

\newcommand{\newtext}[1]{{\textbf{#1}}}
\newcommand{\oldtext}[1]{}

\newcommand{\frbzach}{20220207C}       
\newcommand{\frbalex}{20220307B}       
\newcommand{\frbwhitney}{20220310F}    
\newcommand{\frbmark}{20220319D}       
\newcommand{\frbquincy}{20220418A}     
\newcommand{\frboran}{20220506D}       
\newcommand{\frbjackie}{20220509G}     
\newcommand{\frbansel}{20220825A}      
\newcommand{\frbelektra}{20220914A}    
\newcommand{\frbetienne}{20220920A}    
\newcommand{\frbjuan}{20221012A}       

\begin{document}

\title{Deep Synoptic Array Science: First FRB and Host Galaxy Catalog}

\correspondingauthor{Casey J. Law}
\email{claw@astro.caltech.edu}

\author{Casey J. Law}
\altaffiliation{These authors contributed equally to this work.}
\affiliation{Cahill Center for Astronomy and Astrophysics, MC 249-17 California Institute of Technology, Pasadena CA 91125, USA}
\affiliation{Owens Valley Radio Observatory, California Institute of Technology, Big Pine CA 93513, USA}

\author{Kritti Sharma}
\altaffiliation{These authors contributed equally to this work.}
\affiliation{Cahill Center for Astronomy and Astrophysics, MC 249-17 California Institute of Technology, Pasadena CA 91125, USA}

\author{Vikram Ravi}
\affiliation{Cahill Center for Astronomy and Astrophysics, MC 249-17 California Institute of Technology, Pasadena CA 91125, USA}
\affiliation{Owens Valley Radio Observatory, California Institute of Technology, Big Pine CA 93513, USA}

\author{Ge Chen}
\affiliation{Cahill Center for Astronomy and Astrophysics, MC 249-17 California Institute of Technology, Pasadena CA 91125, USA}

\author{Morgan Catha}
\affiliation{Owens Valley Radio Observatory, California Institute of Technology, Big Pine CA 93513, USA}

\author{Liam Connor}
\affiliation{Cahill Center for Astronomy and Astrophysics, MC 249-17 California Institute of Technology, Pasadena CA 91125, USA}

\author{Jakob T. Faber}
\affiliation{Cahill Center for Astronomy and Astrophysics, MC 249-17 California Institute of Technology, Pasadena CA 91125, USA}

\author{Gregg Hallinan}
\affiliation{Cahill Center for Astronomy and Astrophysics, MC 249-17 California Institute of Technology, Pasadena CA 91125, USA}
\affiliation{Owens Valley Radio Observatory, California Institute of Technology, Big Pine CA 93513, USA}

\author{Charlie Harnach}
\affiliation{Owens Valley Radio Observatory, California Institute of Technology, Big Pine CA 93513, USA}

\author{Greg Hellbourg}
\affiliation{Cahill Center for Astronomy and Astrophysics, MC 249-17 California Institute of Technology, Pasadena CA 91125, USA}
\affiliation{Owens Valley Radio Observatory, California Institute of Technology, Big Pine CA 93513, USA}

\author{Rick Hobbs}
\affiliation{Owens Valley Radio Observatory, California Institute of Technology, Big Pine CA 93513, USA}

\author{David Hodge}
\affiliation{Cahill Center for Astronomy and Astrophysics, MC 249-17 California Institute of Technology, Pasadena CA 91125, USA}

\author{Mark Hodges}
\affiliation{Owens Valley Radio Observatory, California Institute of Technology, Big Pine CA 93513, USA}

\author{James W. Lamb}
\affiliation{Owens Valley Radio Observatory, California Institute of Technology, Big Pine CA 93513, USA}

\author{Paul Rasmussen}
\affiliation{Owens Valley Radio Observatory, California Institute of Technology, Big Pine CA 93513, USA}

\author{Myles B. Sherman}
\affiliation{Cahill Center for Astronomy and Astrophysics, MC 249-17 California Institute of Technology, Pasadena CA 91125, USA}

\author{Jun Shi}
\affiliation{Cahill Center for Astronomy and Astrophysics, MC 249-17 California Institute of Technology, Pasadena CA 91125, USA}

\author{Dana Simard}
\affiliation{Cahill Center for Astronomy and Astrophysics, MC 249-17 California Institute of Technology, Pasadena CA 91125, USA}

\author{Reynier Squillace}
\affiliation{Cahill Center for Astronomy and Astrophysics, MC 249-17 California Institute of Technology, Pasadena CA 91125, USA}
\affiliation{Steward Observatory, University of Arizona, 933 N. Cherry Avenue, Tucson, AZ 85721, USA}

\author{Sander Weinreb}
\affiliation{Cahill Center for Astronomy and Astrophysics, MC 249-17 California Institute of Technology, Pasadena CA 91125, USA}

\author{David P. Woody}
\affiliation{Owens Valley Radio Observatory, California Institute of Technology, Big Pine CA 93513, USA}

\author{Nitika Yadlapalli}
\affiliation{Cahill Center for Astronomy and Astrophysics, MC 249-17 California Institute of Technology, Pasadena CA 91125, USA}

\begin{abstract}

Fast Radio Bursts (FRBs) are a powerful and mysterious new class of transient that are luminous enough to be detected at cosmological distances. By associating FRBs to host galaxies, we can measure intrinsic and environmental properties that test FRB origin models, in addition to using them as precise probes of distant cosmic gas. The \oldtext{110-antenna} Deep Synoptic Array (DSA-110) is a radio interferometer built to maximize the rate at which it can simultaneously detect and localize FRBs. Here, we present the first sample of FRBs and host galaxies discovered by the DSA-110. This sample of 11 FRBs is the largest\newtext{, most} uniform sample of localized FRBs to date \oldtext{and}\newtext{, as it} is selected based on association to host galaxies identified in optical imaging by Pan-STARRS1 \oldtext{and follow-up spectroscopy at the Palomar and Keck observatories}. These FRBs have not been observed to repeat and their radio properties (dispersion, temporal scattering, energy) are similar to that of the known non-repeating FRB population. Most host galaxies have ongoing star formation, as has been identified before for FRB hosts. \oldtext{In contrast to prior work, a large fraction (four of eleven)}\newtext{Two hosts} of the new sample are \oldtext{more} massive\newtext{, quiescent galaxies.} \oldtext{than 10$^{11}$\ M$_{\odot}$ and most had elevated star formation rates more than 100 Myr in their past.}
The distribution of star-formation history across this host-galaxy sample shows that the delay-time distribution is wide, \newtext{with a powerlaw model that spans} \oldtext{spanning} from $\sim100$\,Myr to \oldtext{$\sim10$\,}\newtext{$\gtrsim2$}Gyr. This requires the existence of one or more progenitor formation channels associated with old stellar populations, such as the binary evolution of compact objects.
\end{abstract}


\section{Introduction} \label{sec:intro}

Fast Radio Bursts (FRBs) are a new class of $\ll1$\,s radio transient that is powerful enough to be seen from distant galaxies \citep{2019ARA&A..57..417C,2022A&ARv..30....2P}. The nature of the source (or sources) that make FRBs is not known yet. Several hundred bursts have been detected \citep{2021ApJS..257...59C}, some of which repeat stochastically \citep{2023arXiv230108762T} and some of which burst with quasiperiodic patterns over a wide range of timescales \citep{2020Natur.582..351C,arXiv:2202.08002}. Polarimetric measurements have identified dynamic, highly magnetized plasma in the parsec-scale environments of some FRB sources \citep{2022arXiv220211112A,2023arXiv230208386M}.

The need to localize FRBs to an arcsecond (or better) precision has motivated a new generation of radio instruments \citep[e.g.,][]{2010PASA...27..272M,2018ApJS..236....8L,2018IAUS..337..406S,2023A&A...672A.117V}. Of the roughly two dozen localized FRBs, the story told by their host galaxies and host environments is confusing. Most FRBs are associated with actively star-forming galaxies \citep{2017Natur.541...58C,2021ApJ...917...75M,2022AJ....163...69B}, consistent with the discovery of FRB-like bursts from a magnetar in our own galactic disk \citep{2020Natur.587...59B,2020Natur.587...54C}. However, some FRBs have been associated with much older stellar systems, including a nearby globular cluster~\citep{2022Natur.602..585K} and a quiescent galaxy \citep{2023arXiv230214782S}. A small fraction have a \newtext{compact,} persistent radio \newtext{source (PRS), that is coincident with the FRB}, but many other FRBs have strict limits on \newtext{such}\oldtext{associated persistent} emission \citep{2022Natur.606..873N,2020Natur.577..190M,2022ApJ...927...55L}. Therefore, a concordant picture of the FRB progenitors has not yet emerged from studies of FRB hosts. 

A major question raised with the discovery of FRBs remains unanswered: what source (or sources) produces these powerful bursts? Their occurrence rate \citep[$\sim10^5$\ Gpc$^{-3}$\ yr$^{-1}$ for burst energy $>10^{39}$\ erg/Hz;][]{2019NatAs...3..928R,2022MNRAS.510L..18J} is too high to be attributed solely to cataclysmic events such as classes of supernova. Models based on magnetized neutron stars can reproduce many burst properties and FRB associations \citep{2020Natur.587...45Z}. However, even this one kind of object can be formed in young (through core-collapse supernovae) or older (through compact binary coalescence) stellar environments \citep{2019ApJ...886..110M,2021ApJ...917L..11K}. Some FRB behaviors, such as quasiperiodic activity timescales, have been used to argue that binarity plays a role \citep{2020Natur.582..351C,2020ApJ...893L..39L,2022arXiv220507917L}. Finally, as proposed in \citet{2023arXiv230205465G} and \citet{2023arXiv230214782S}, multiple formation channels may exist and be associated with specific stellar environments, as is the case for supernovae \citep{2020ApJ...904...35P,2022MNRAS.510.1867L} and gamma-ray bursts \citep{2014ARA&A..52...43B,2006ARA&A..44..507W}. If so, then a large sample of FRB host galaxy associations and careful statistical analysis will be required to identify specific classes of FRB source and their formation channels.

Here, we present the first sample of FRBs discovered by the DSA-110. This sample was selected from FRB discoveries made during science commissioning from February to October 2022. We selected all FRBs that are associated with a host galaxy in Pan-STARRS1, which we used to guide follow-up optical spectroscopic and infrared photometric observations. This sample represents a significant increase in the number of FRBs with quality host galaxies identified over the last seven years \citep[e.g.,][]{2017ApJ...834L...7T,2020Natur.577..190M,2022MNRAS.514.1961R, 2022AJ....163...69B}. Three FRBs included in this sample have been reported in other publications \citep[FRBs \frbmark, \frbjackie, \frbelektra;][]{2023ApJ...949L...3R, arXiv:2302.14788, 2023arXiv230214782S} and this work uses the same instrument and analysis techniques described previously.

Our goal is to present a \oldtext{uniform}\newtext{host magnitude limited} sample of FRBs and their host galaxies to probe ideas for FRB origin (see \S \ref{sec:obs}). The DSA-110 was in science commissioning during this observing period, so the overall observing efficiency is not measured well enough to estimate an FRB rate. However, the FRB characterization and host identification is reliable enough to use this sample to study burst phenomenology and galaxy properties (see \S \ref{sec:ana}). In \S \ref{sec:dis}, we discuss details of FRB host galaxies and how their star-formation history is most consistent with FRB sources formed over a wide range of delay times. This is supported by analysis of the nearest FRBs and hosts, for which we have more reliable host identification and high physical resolution. We conclude by discussing FRB origins and whether multiple models are required in \S \ref{sec:con}. We adopt standard cosmological parameters from \citet{Planck18}.

\section{Observations}
\label{sec:obs}

\subsection{DSA-110}

The DSA-110 is built to discover, localize, and characterize FRBs. The first FRB localization by the instrument demonstrated the techniques used for the current sample \citep{2023ApJ...949L...3R}. A full description of the DSA-110 interferometer and FRB search system will be presented in Ravi et al (in prep). 

FRBs described here were detected during science commissioning with an array of 63, 4.65-meter antennas distributed over a 2.5-km area at the Owens Valley Radio Observatory\footnote{See \url{https://www.ovro.caltech.edu}.}. The FRB search system uses 48 antennas arranged as a linear array to form 256 coherent ``fan beams'' spaced by 1\arcmin, thus spanning a little beyond the primary beam width of 3.4 deg (FWHM). The DSA-110 is a transit telescope with motorized control over antenna elevation. During all observations described here, the array was pointed at a declination of 71.6$^{\circ}$. 
 
The digital and software systems search a continual stream of calibrated power beams in real time for FRBs. The FRB search is designed to avoid and remove radio frequency interference (RFI) by pointing away from known RFI sources, automatically suspending search during RFI ``storms'', and data cleaning. The search system is optimally sensitive to temporal widths from 0.26\,ms to 8.32\,ms, and FRB-like dispersion measures (DM) of 50\,pc\,cm$^{-3}$ to 1500\,pc\,cm$^{-3}$. For the fixed declination of the observations, the instrument observed Galactic latitudes from 8$^{\circ}$\ to 46$^{\circ}$, which corresponds to a Milky Way DM contribution ranging from 133 to 36 pc cm$^{-3}$ \citep{2002astro.ph..7156C}.

The search system is designed to automatically identify FRBs based on several criteria. The significance of power excess in any beam must be greater than 8.5 and not associated with strong terrestrial interference, and have a DM greater than both 50\,pc\,cm$^{-3}$ and 0.75 times the Galactic DM contribution. For each candidate FRB, the system saves metadata, and channelized voltage data from the entire array (48 search antennas, plus 15 outrigger antennas). The voltage data can fully reproduce the real-time search data, as well as enable fine localization, high time- and frequency-resolution, and full polarimetry. Verification of each candidate is ultimately reliant on a successful interferometric localization. 

\begin{figure*}
    \centering
    \includegraphics[width=0.3\textwidth]{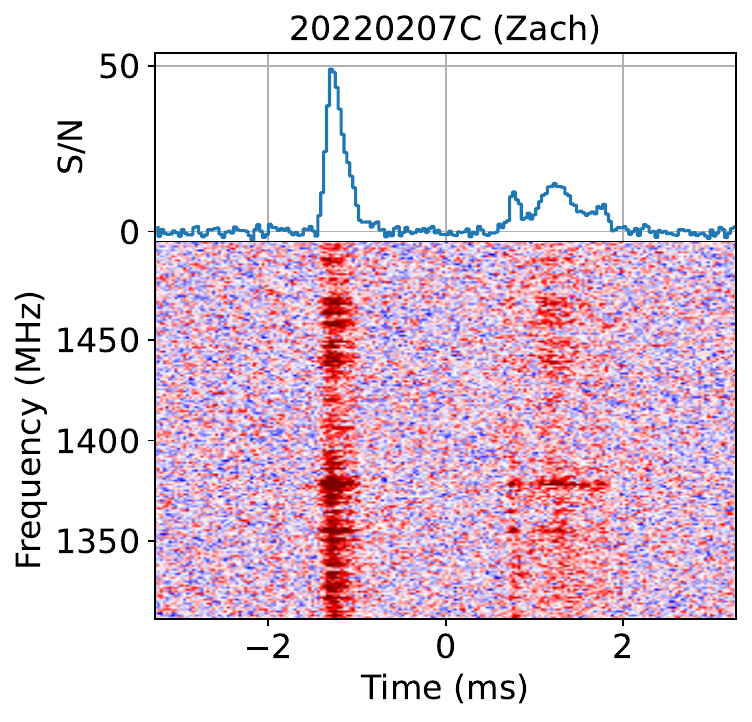}
    \includegraphics[width=0.3\textwidth]{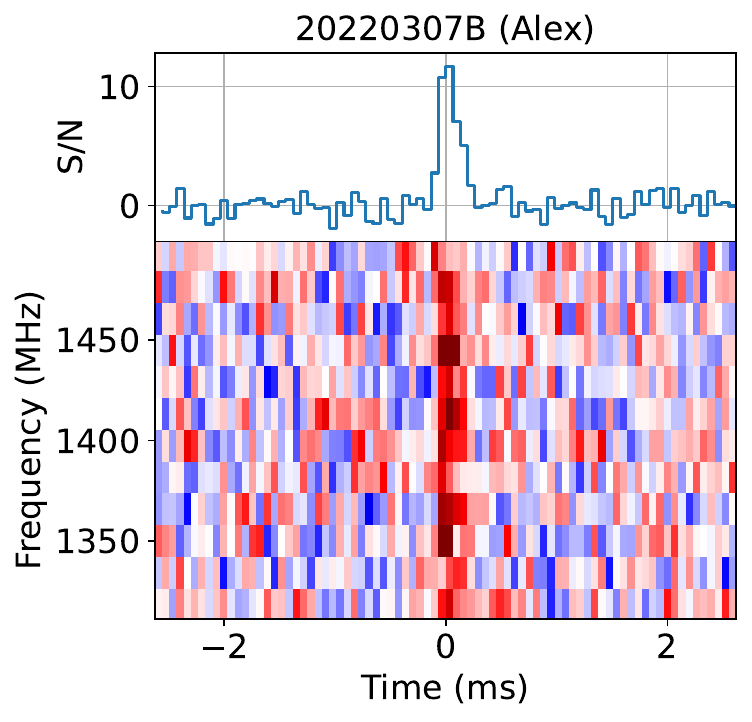}
    \includegraphics[width=0.3\textwidth]{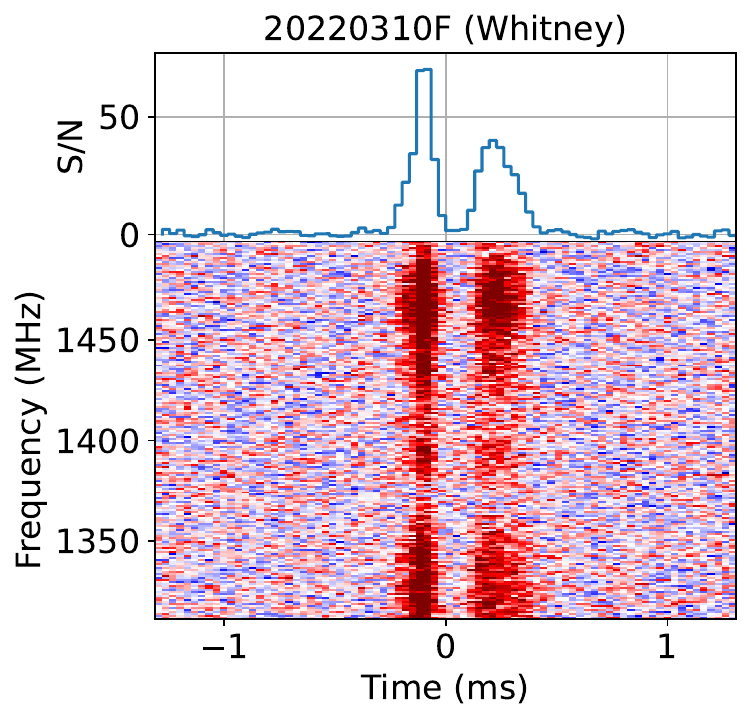}

    \includegraphics[width=0.3\textwidth]{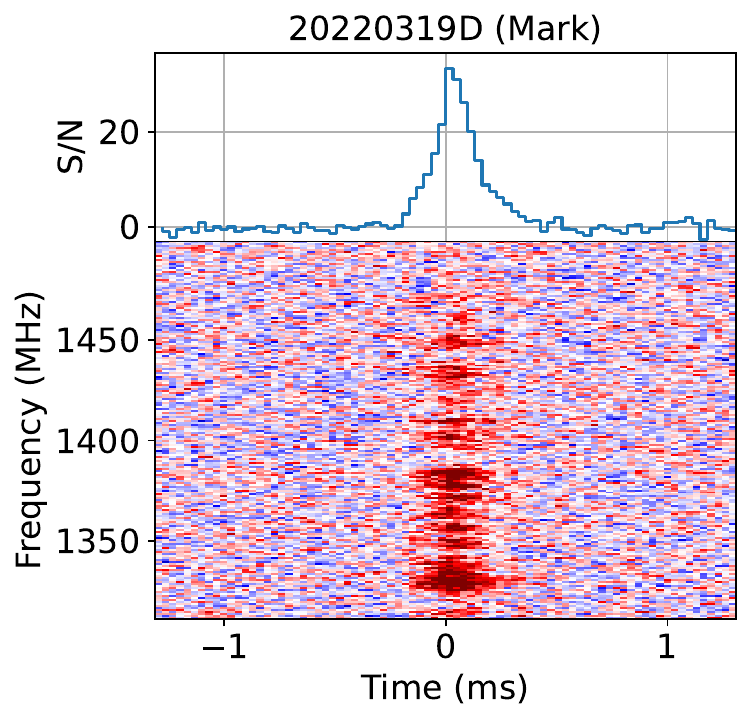}
    \includegraphics[width=0.3\textwidth]{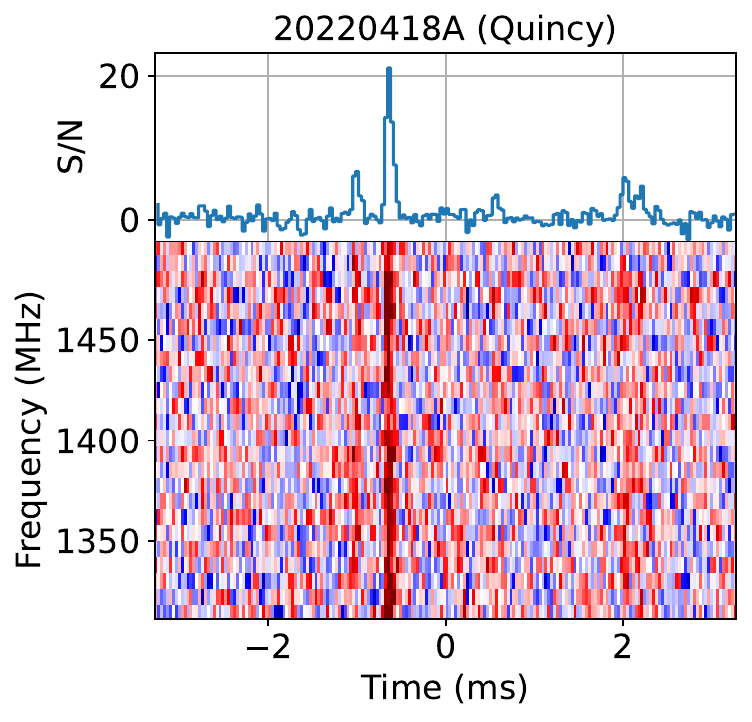}
    \includegraphics[width=0.3\textwidth]{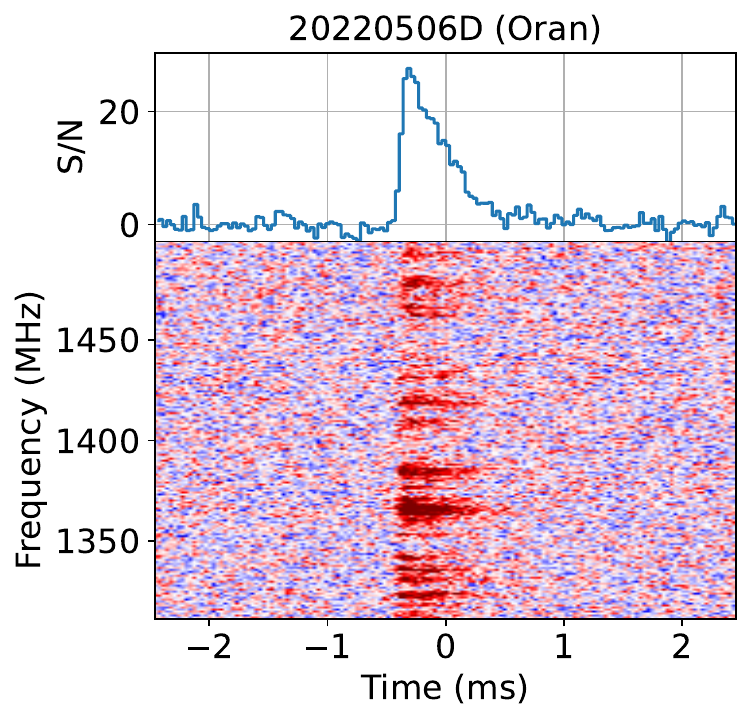}

    \includegraphics[width=0.3\textwidth]{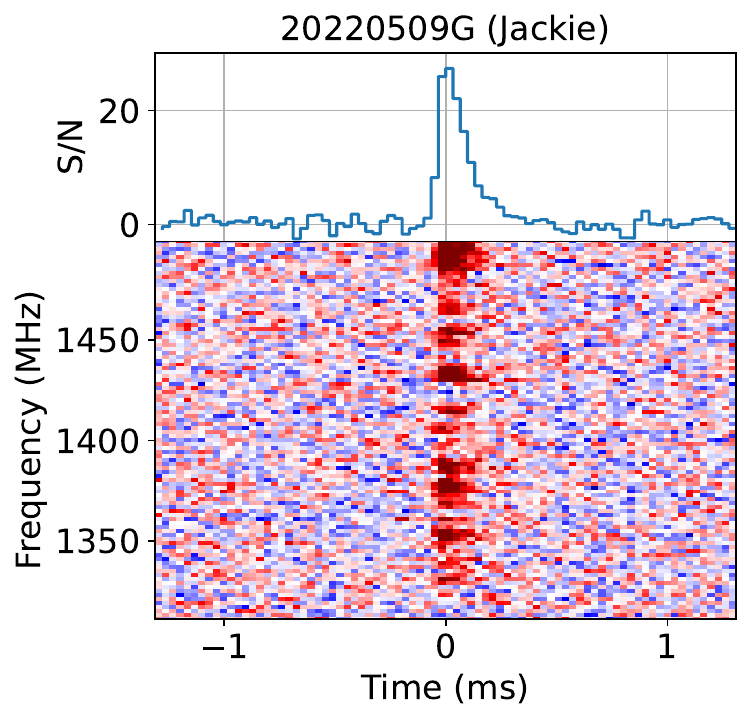}
    \includegraphics[width=0.3\textwidth]{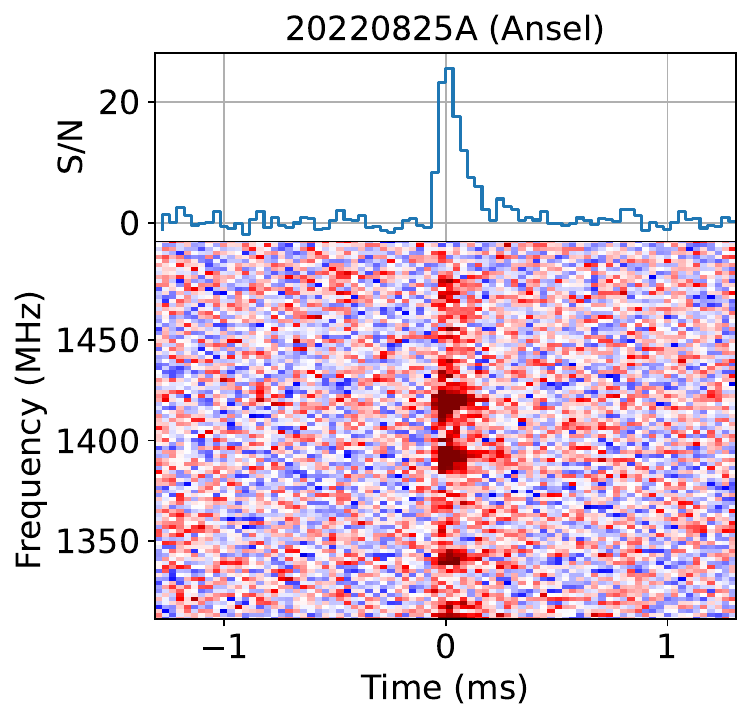}
    \includegraphics[width=0.3\textwidth]{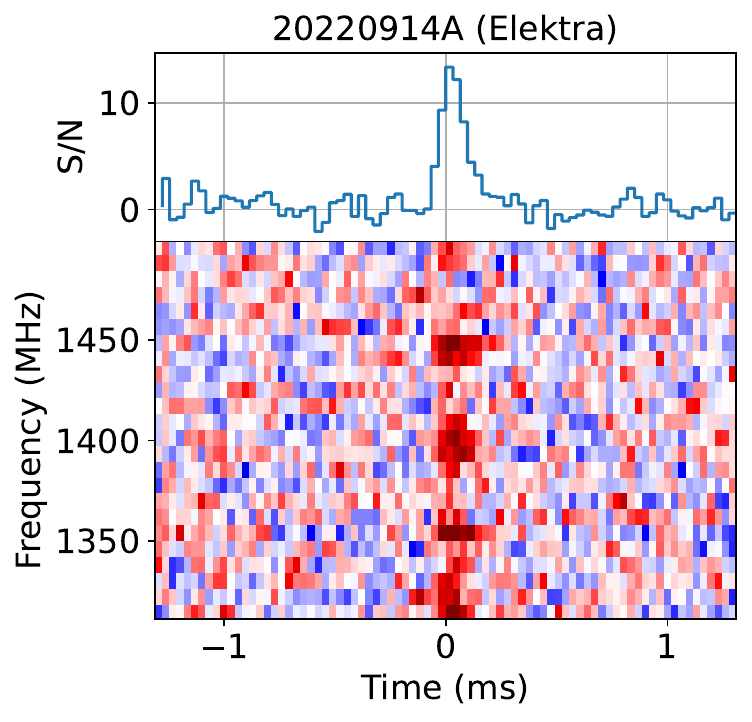}

    \includegraphics[width=0.3\textwidth]{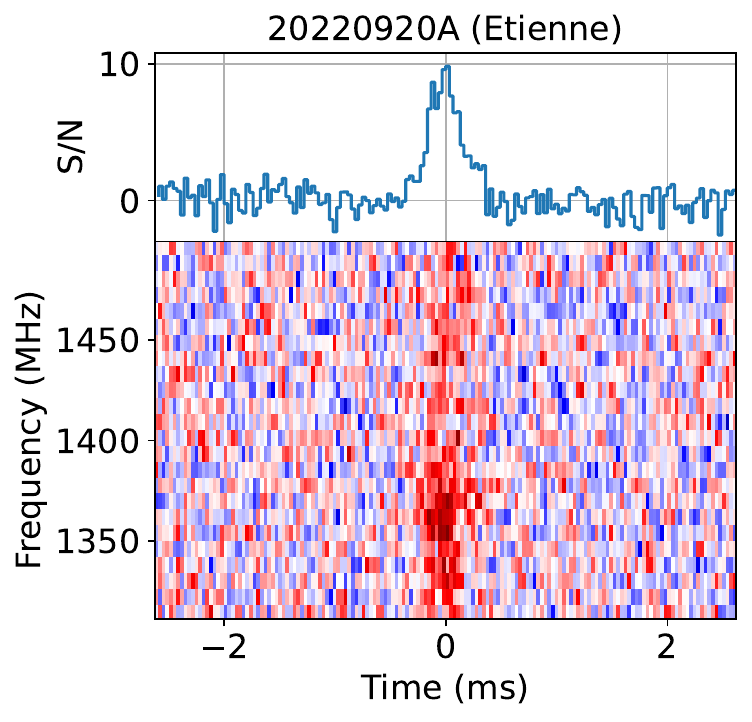}
    \includegraphics[width=0.3\textwidth]{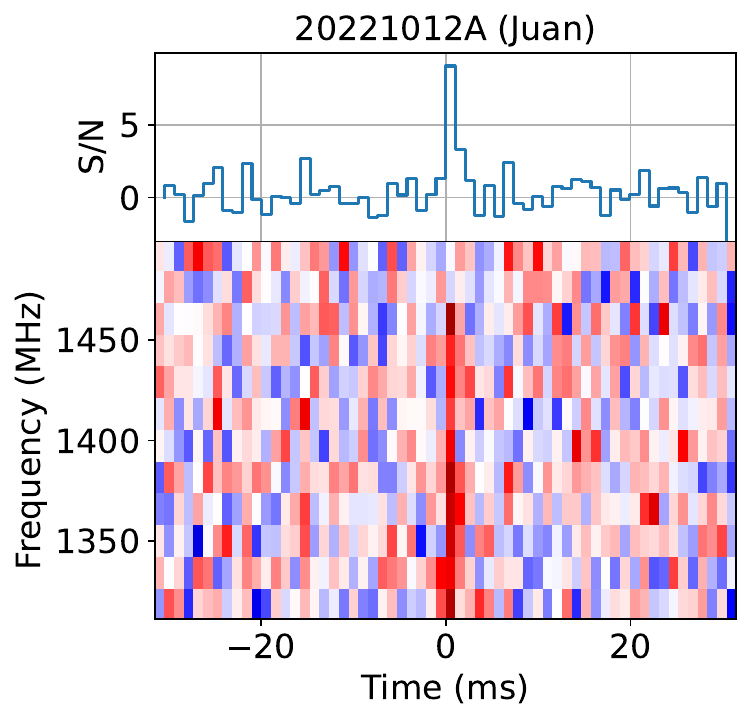}

    \caption{Dynamic spectra for DSA-110 FRB discoveries, listed in discovery order. The colormap shows the power in the tied-array beam, while the top panel shows the frequency-averaged light curve in units of SNR. (Top row) Plots for FRBs \frbzach, \frbalex, \frbwhitney. (Second row) Plots for FRBs \frbmark, \frbquincy, \frboran. (Third row) Plots for FRBs \frbjackie, \frbansel, \frbelektra. (Bottom Row) Plots for FRBs \frbetienne, \frbjuan.}
    \label{fig:bursts}
\end{figure*}

Table \ref{tab:frbs} and Figure \ref{fig:bursts} present the first sample of 11 FRBs detected by DSA-110 and robustly associated with counterparts detected by Pan-STARRS1 (\S \ref{sec:assoc}). These events were identified during science commissioning from January to October 2022. Analysis of this sample is presented in \S \ref{sec:bursts}. A more detailed analysis of the burst properties, including full polarimetry, \oldtext{will be}\newtext{is} presented in \newtext{\citet{2023arXiv230806813S}, \citet{2023arXiv230806816S}}\oldtext{Sherman et al. (in prep)} and Chen et al (in prep). DSA-110 FRB detections, including data to aid in reproducing this analysis, are available on the \dataset[DSA-110 archive]{http://code.deepsynoptic.org/dsa110-archive} and in the FRB software and data repository \citep{10.5281/zenodo.8125230}.

\begin{table*}[ht]
\begin{tabular}{l|ccccccc}
FRB Name & Time & Location & S/N & Fluence & DM & DM$_{\rm{ISM}}$ & Width \\ 
 & (MJD) & (J2000) & & (Jy ms) & (pc/cm3) & (pc/cm3) & (ms) \\ \hline
\frbzach\ / ``Zach'' & 59617.808504 & 20:40:47.886, +72:52:56.378 & 60.0 & 16.2 & 262.38 & 79.3 & 0.5 \\ 
 & & (0.63, 0.54) & & & 0.01 & & \\
\frbalex\ / ``Alex'' & 59645.845634 & 23:23:29.88, +72:11:32.6 & 11.9 & 3.2 & 499.27 & 135.7 & 0.5 \\ 
 & & (1.72, 1.26) & & & 0.06 & & \\
\frbwhitney\ / ``Whitney'' & 59648.241721 & 8:58:52.9, +73:29:27.0 & 68.4 & 26.2 & 462.24 & 45.4 & 1.0 \\ 
 & & (1.9, 1.4) & & & 0.005 & & \\
\frbmark\ / ``Mark'' & 59657.932757 & 02:08:42.70, +71:02:06.9 & 41.7 & 8.0 & 110.98 & 133.3 & 0.3 \\ 
 & & (0.58, 0.55) & & & 0.02 & & \\
\frbmark\ / ``Mark'' & 59657.932757 & 02:08:42.70, +71:02:06.9 & 41.7 & 8.0 & 110.98 & 133.3 & 0.3 \\ 
 & & (0.58, 0.55) & & & 0.02 & & \\
\frbquincy\ / ``Quincy'' & 59687.355728 & 14:36:25.34, +70:05:45.4 & 10.9 & 4.2 & 623.25 & 37.6 & 1.0 \\ 
 & & (1.356, 0.761) & & & 0.01 & & \\
\frboran\ / ``Oran'' & 59705.597013 & 21:12:10.76, +72:49:38.2 & 48.9 & 13.2 & 396.97 & 89.1 & 0.5 \\ 
 & & (1.11, 0.81) & & & 0.02 & & \\
\frbjackie\ / ``Jackie'' & 59708.4944990805 & 18:50:40.8, +70:14:37.8 & 21.5 & 5.8 & 269.53 & 55.2 & 0.5 \\ 
 & & (2.2, 1.5) & & & 0.02 & & \\
\frbansel\ / ``Ansel'' & 59816.2572924694 & 20:47:55.55, +72:35:05.9 & 15.1 & 5.8 & 651.24 & 79.7 & 1.0 \\ 
 & & (0.78, 0.69) & & & 0.06 & & \\
\frbelektra\ / ``Elektra'' & 59836.145966 & 18:48:13.63, +73:20:12.9 & 9.6 & 2.6 & 631.28 & 55.2 & 0.5 \\ 
 & & (0.93, 0.73) & & & 0.04 & & \\
\frbetienne\ / ``Etienne'' & 59842.9815705163 & 16:01:01.70, +70:55:07.7 & 14.4 & 3.9 & 314.99 & 40.3 & 0.5 \\ 
 & & (1.02, 0.60) & & & 0.01 & & \\
\frbjuan\ / ``Juan'' & 59864.0510912293 & 18:43:11.69, +70:31:27.2 & 9.4 & 5.1 & 441.08 & 54.4 & 2.0 \\ 
 & & (1.12, 0.74) & & & 0.7 & & \\
\end{tabular}
\caption{The first sample of DSA-110 FRBs, listed in discovery order with both formal and informal names in the first column. Time is topocentric and referenced to the top of the observing band at 1530\,MHz. Position errors are listed on alternate rows. Location error is quoted as 1$\sigma$ uncertainties (statistical + systematic). The DM and width maximize S/N of the brightest component; width corresponds to a temporal boxcar size. Fluence is measured from high-resolution data to contain 90\% of the flux density. The DM$_{\rm{ISM}}$ is measured from the model of \citet{2002astro.ph..7156C}. } 
\label{tab:frbs}
\end{table*}

\subsection{Pan-STARRS1}
\label{sec:ps1}

\begin{figure*}
    \centering
    \includegraphics[width=0.3\textwidth]{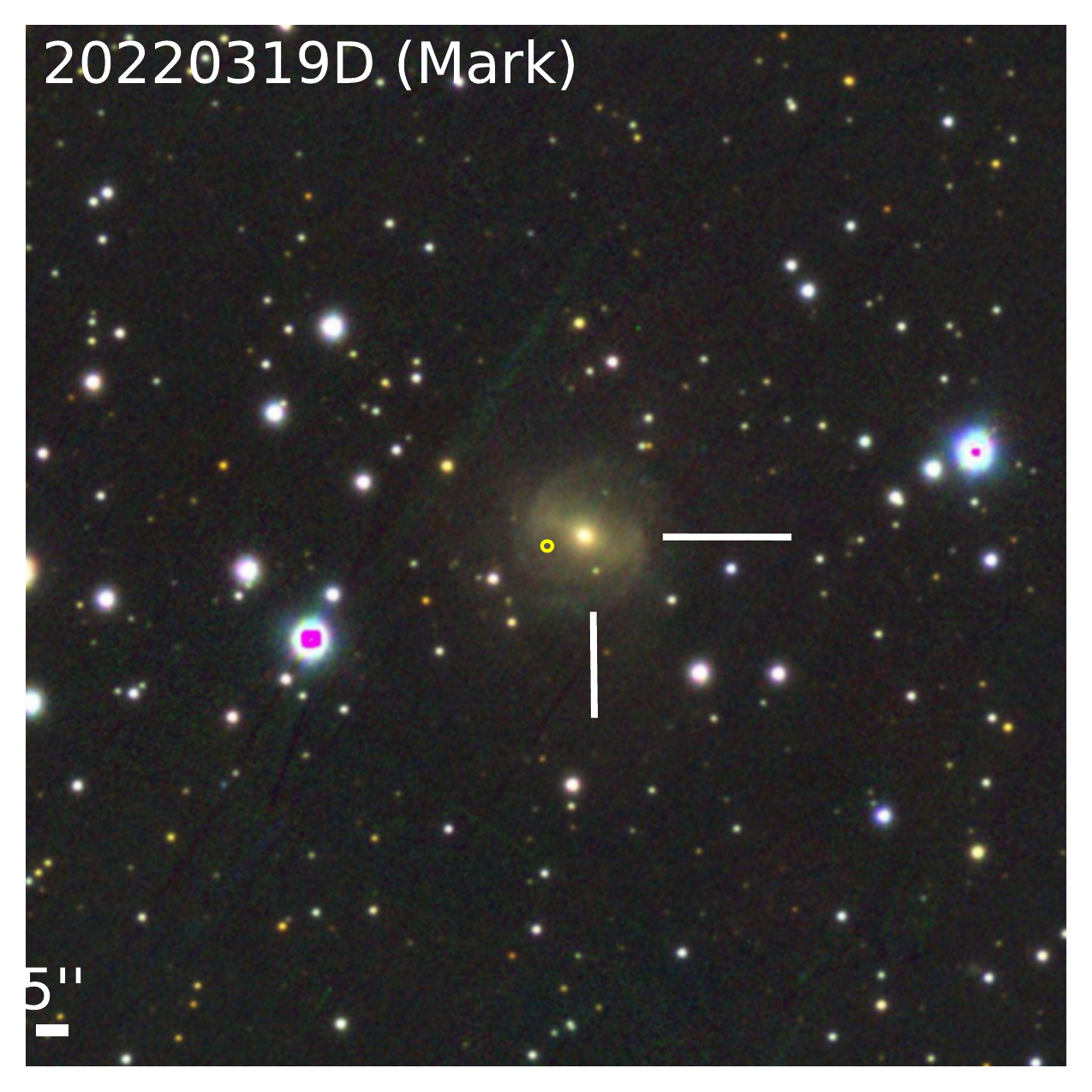}
    \includegraphics[width=0.3\textwidth]{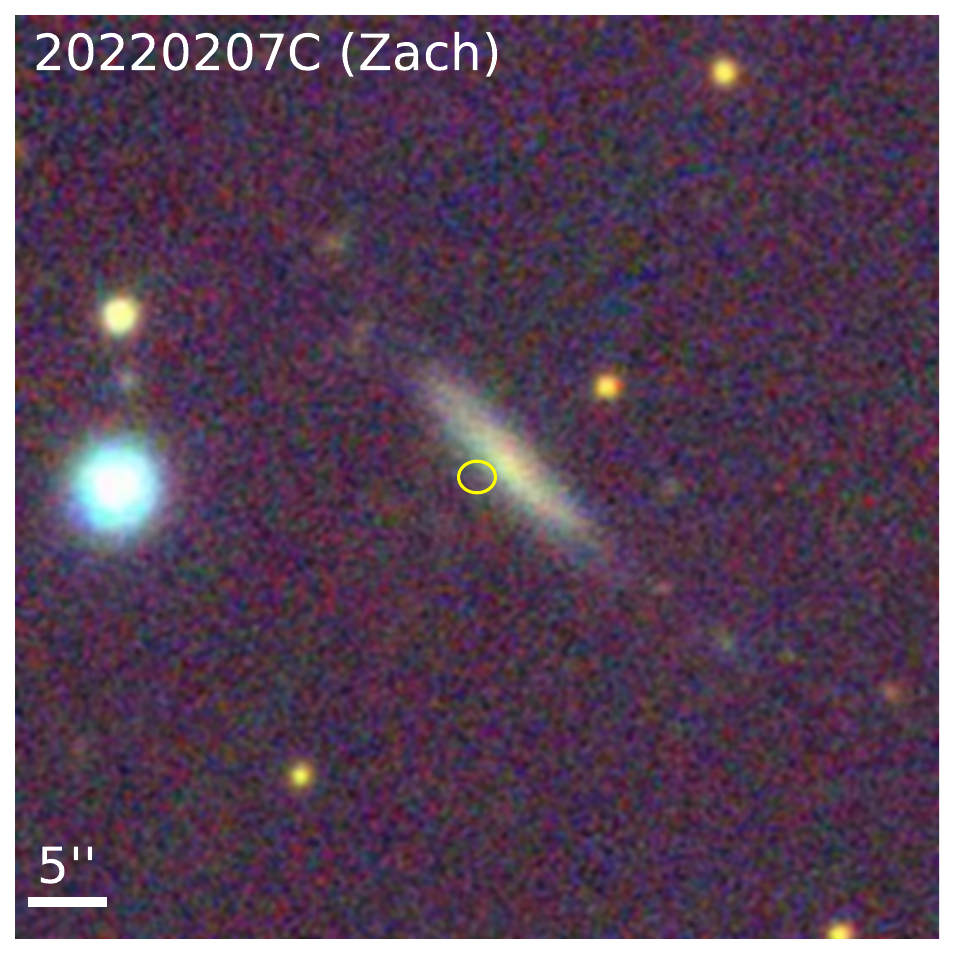}
    \includegraphics[width=0.3\textwidth]{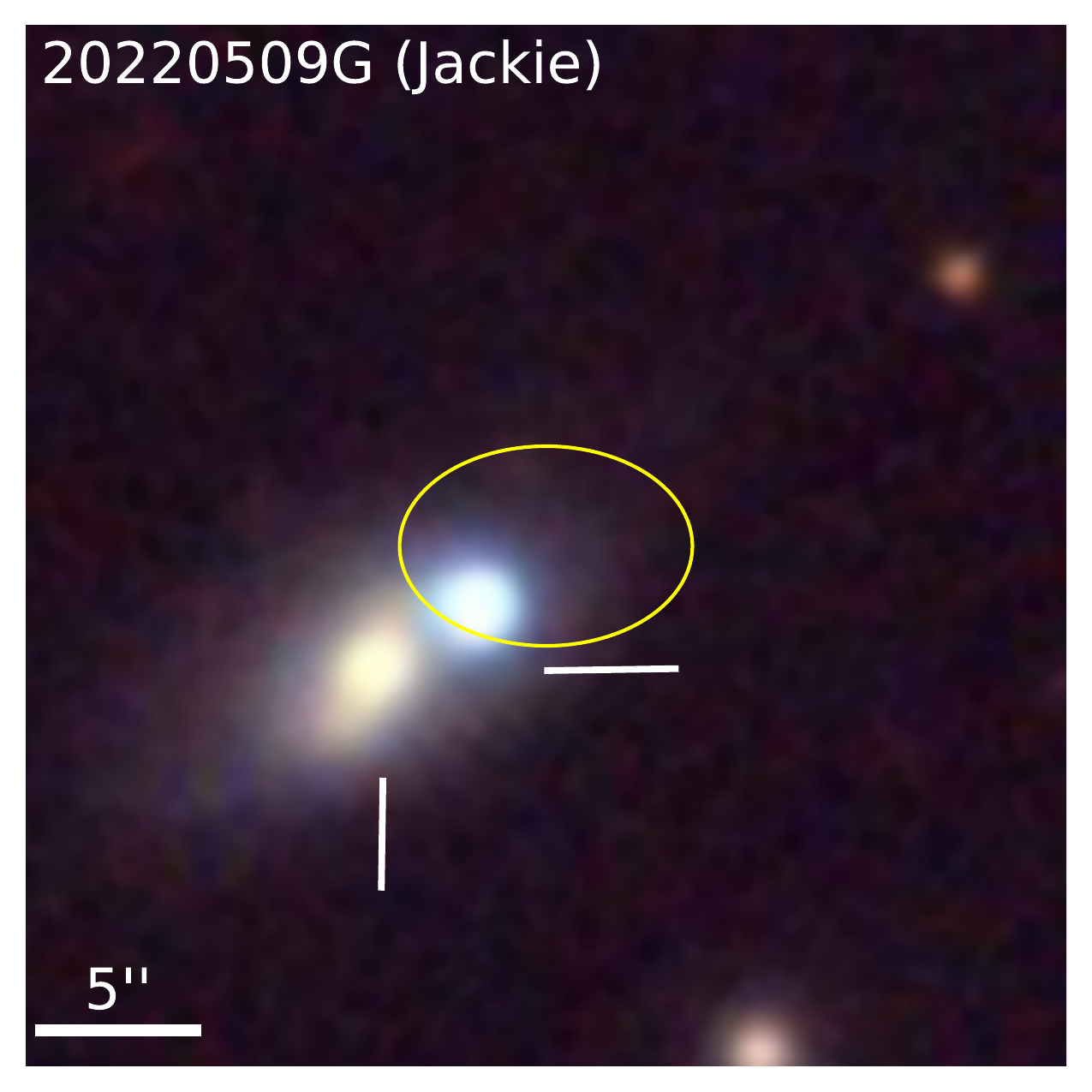}
    
    \includegraphics[width=0.3\textwidth]{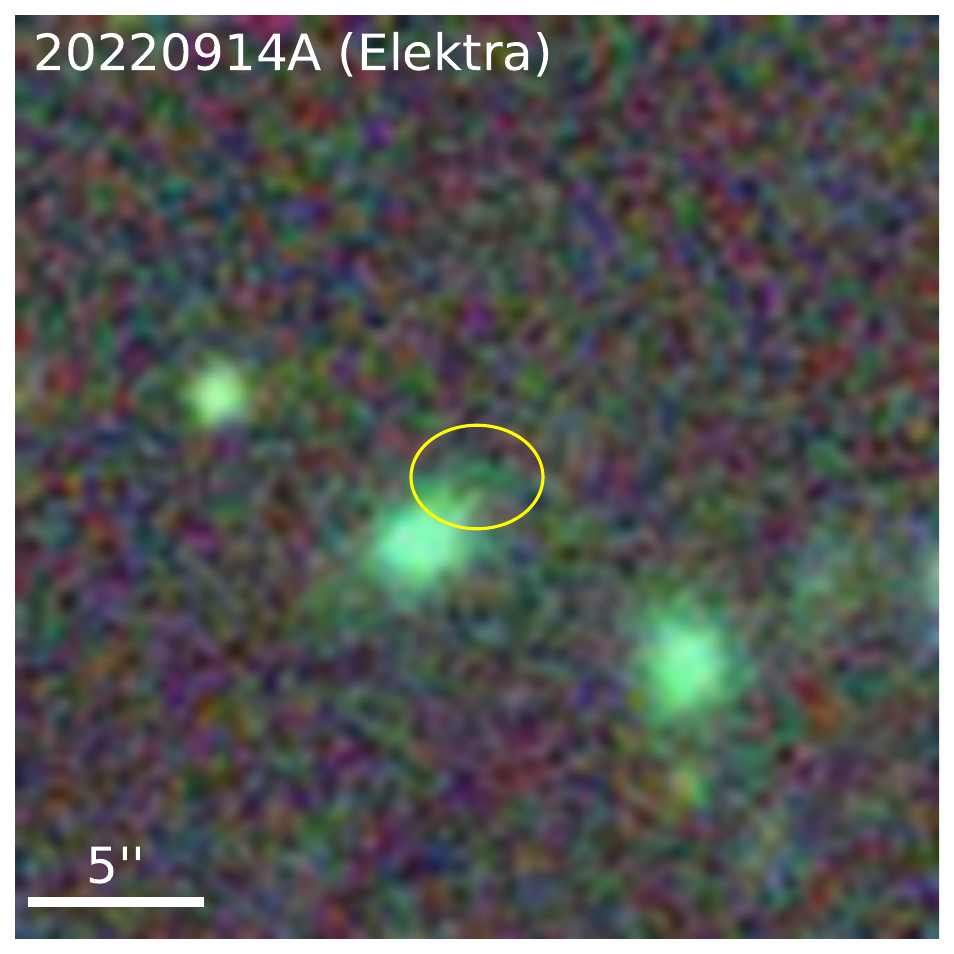}
    \includegraphics[width=0.3\textwidth]{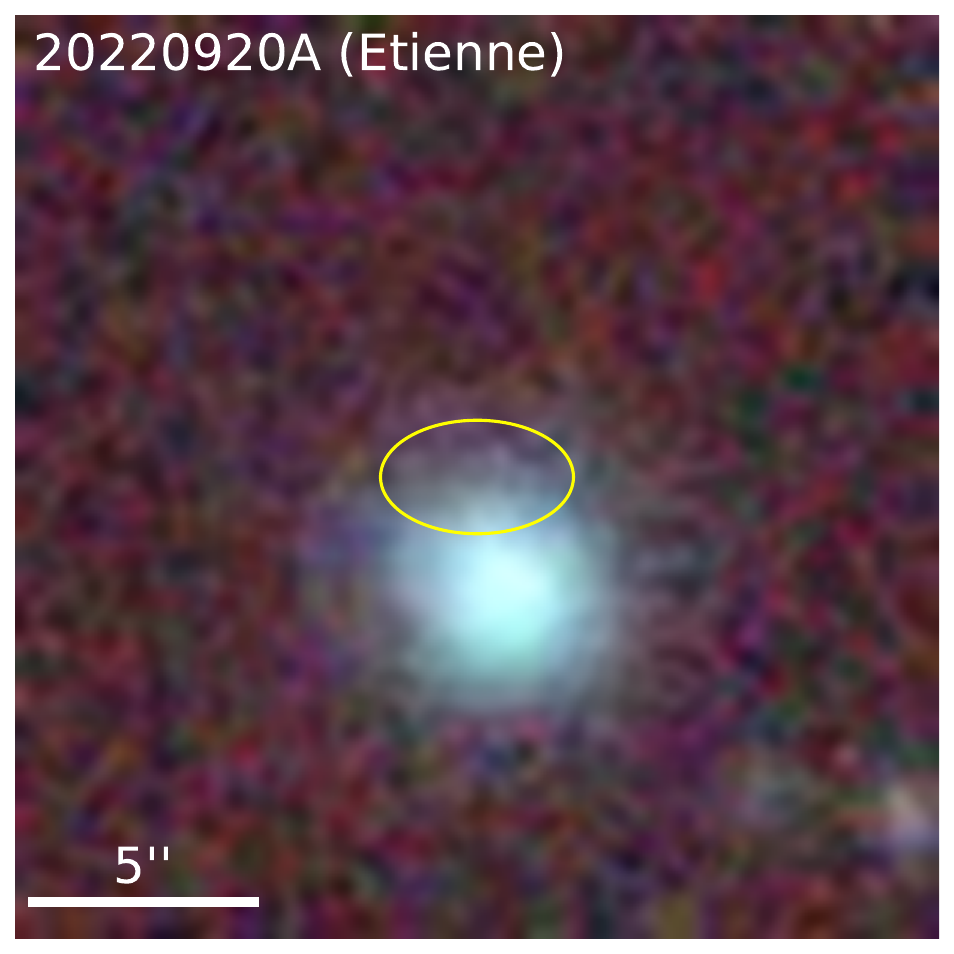}
    \includegraphics[width=0.3\textwidth]{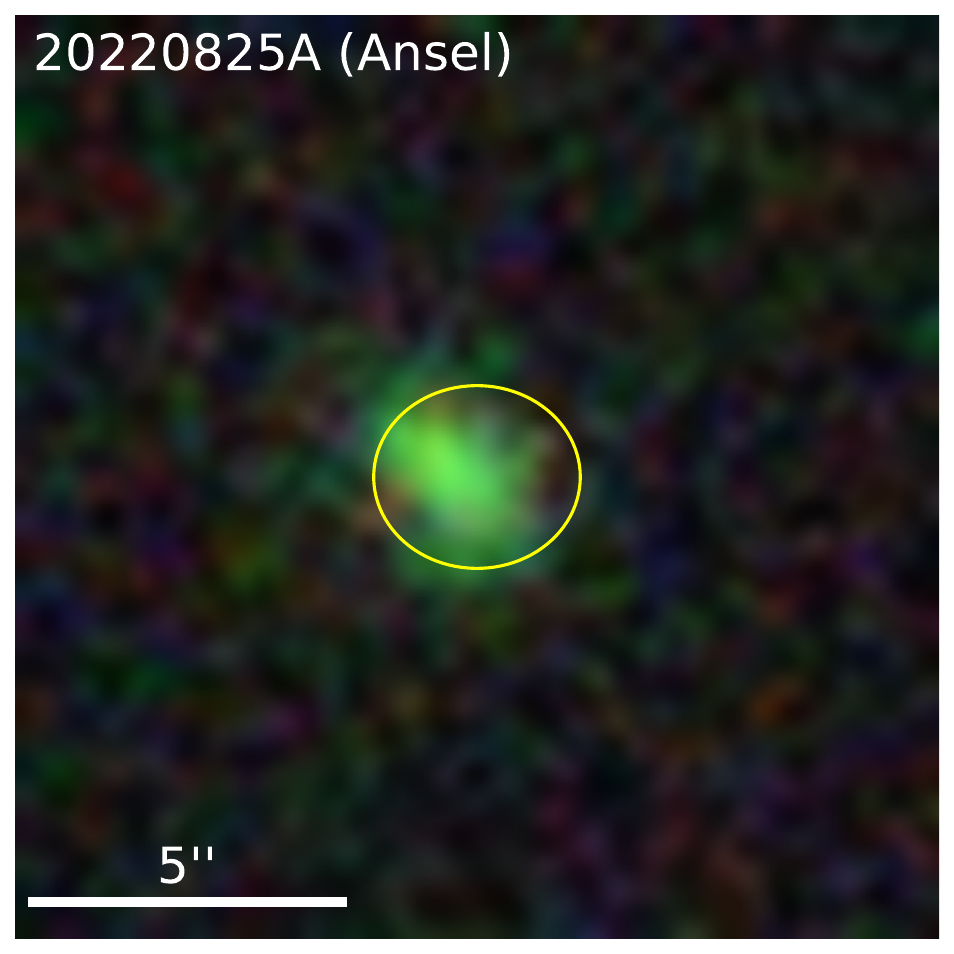}

    \includegraphics[width=0.3\textwidth]{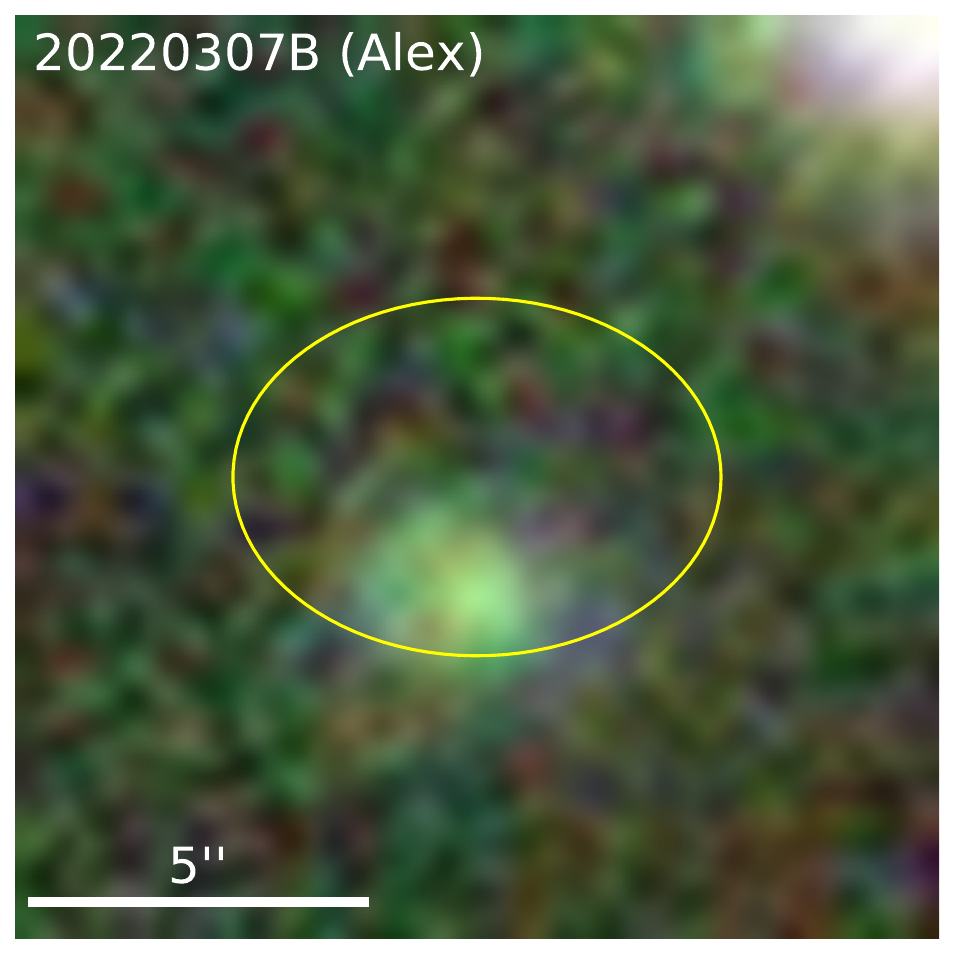}
    \includegraphics[width=0.3\textwidth]{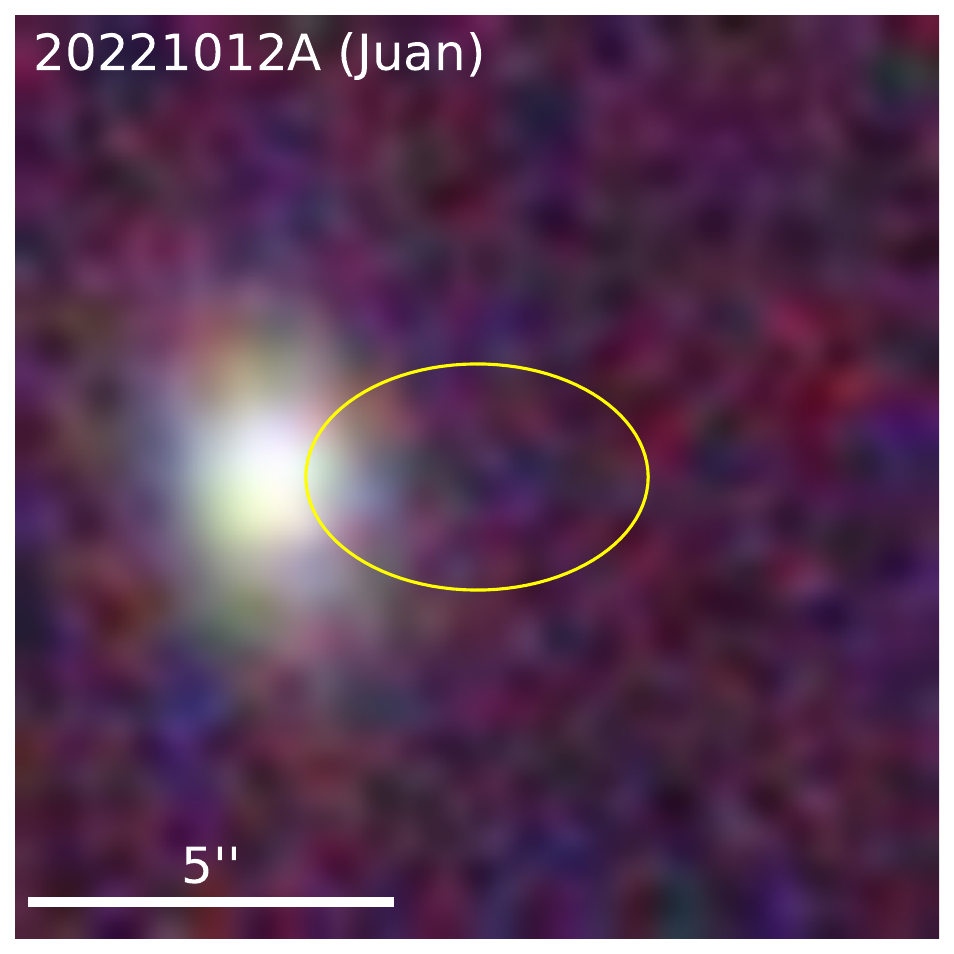}
    \includegraphics[width=0.3\textwidth]{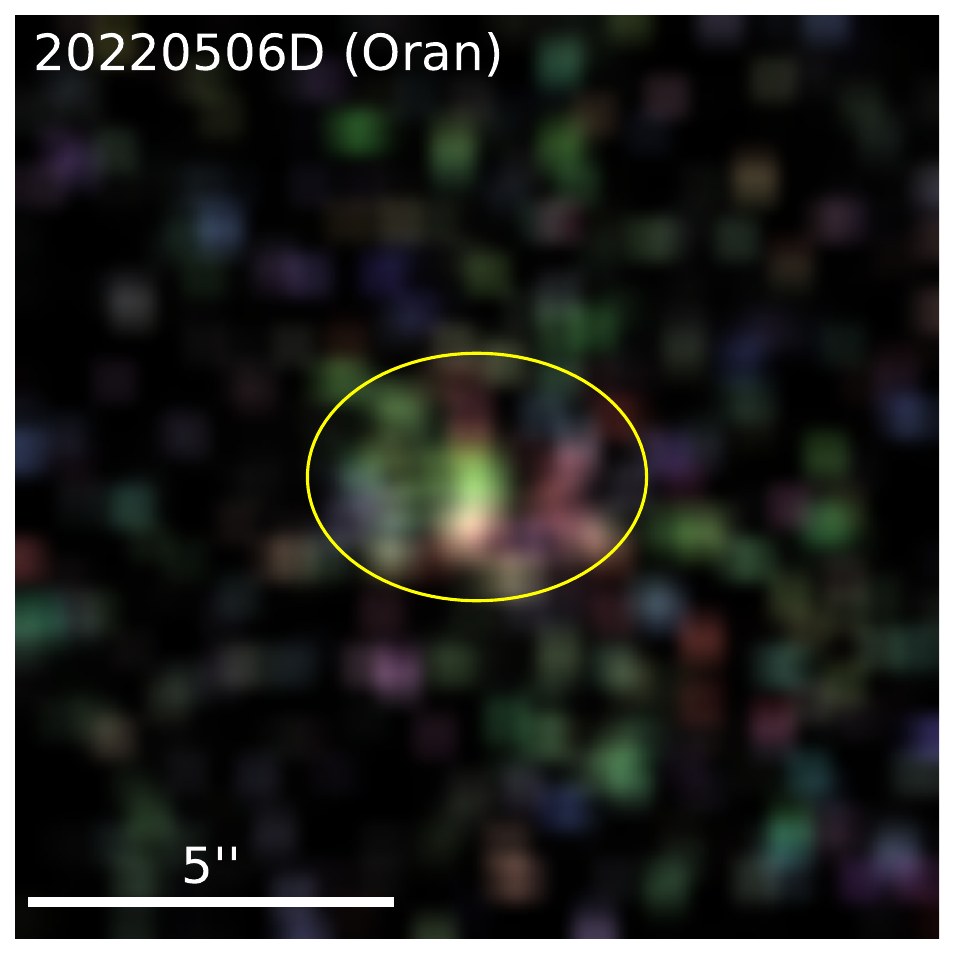}

    \includegraphics[width=0.3\textwidth]{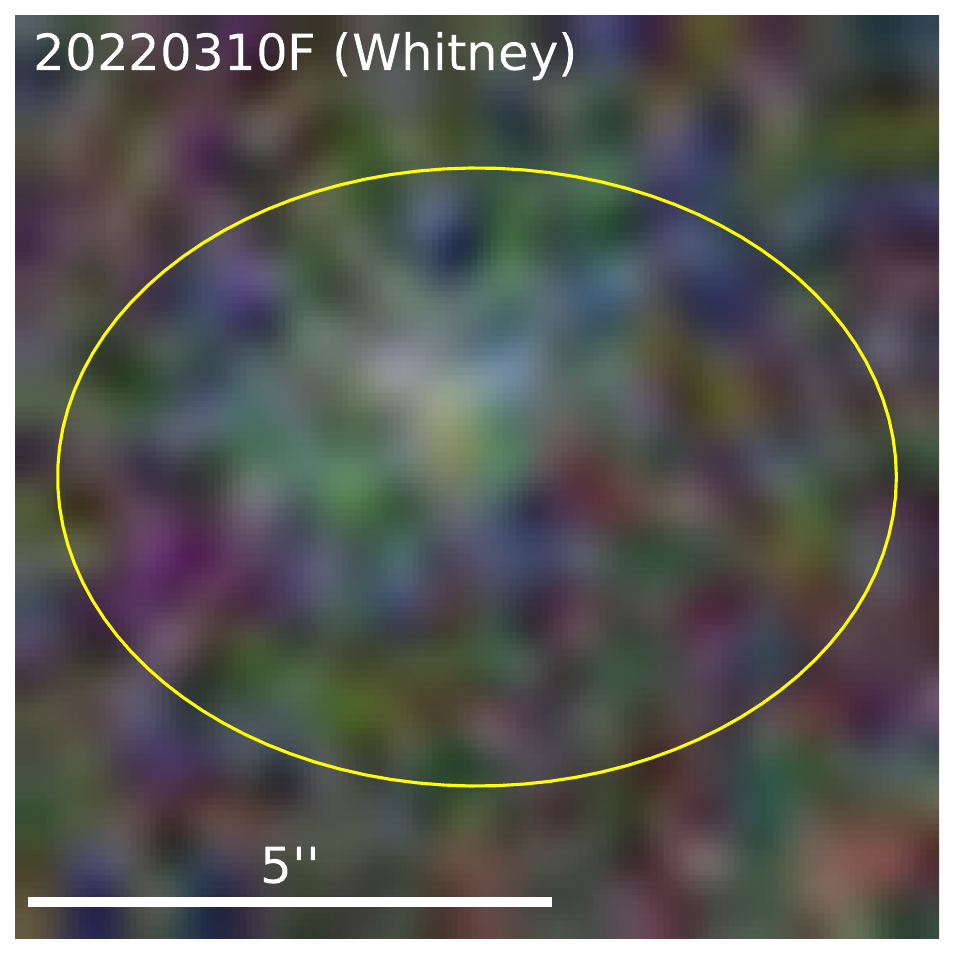}
    \includegraphics[width=0.3\textwidth]{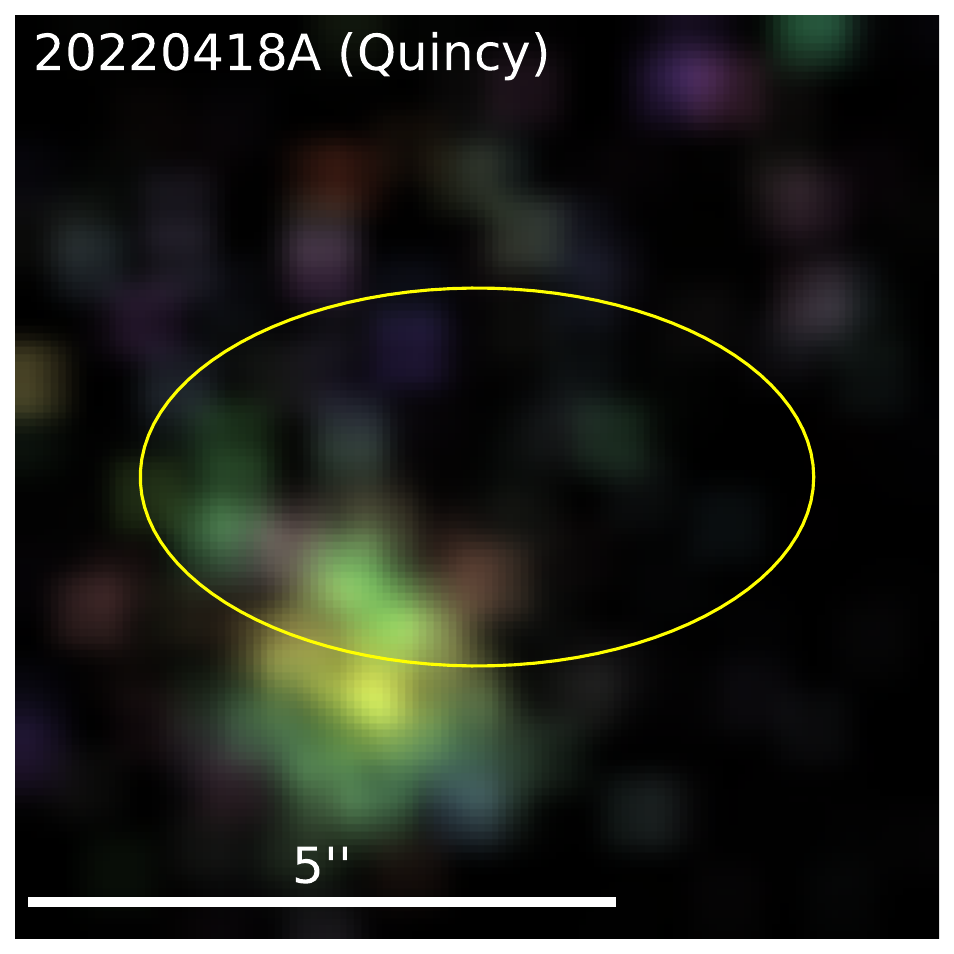}
    \caption{Fixed-scale Pan-STARRS1 images of FRB host galaxies in distance order (see Table \ref{tab:hostderived}). Each image is scaled to span a fixed horizontal size of 60 kpc. The yellow circle shows the FRB localization uncertainty region (90 \% confidence). (Top row) Plots for FRBs \frbmark, \frbzach, \frbjackie. For two of these, we identify the host galaxy with cross hairs. (Second row) Plots for FRBs \frbelektra, \frbetienne, \frbansel. (Third row) Plots for FRBs \frbalex, \frbjuan, \frboran. (Bottom row) Plots for FRBs \frbwhitney, \frbquincy.}
    \label{fig:host}
\end{figure*}

Figure \ref{fig:host} shows Pan-STARRS1 \citep[PS1;][]{2016arXiv161205560C} $gri$ imaging of the likely FRB host galaxies. PS1 provides images and catalogs for the entire sky north of a declination of --30 deg (the ``3pi Steradian Survey''). Multi-epoch image stacks reach a depth of 21 to 23rd magnitude (5$\sigma$) with a median seeing from 1.3 to 1.0 arcsec in $g$, $r$, $i$, $z$, and $y$ bands.

DSA-110 localizes FRBs with a precision better than roughly $\pm2$ arcsec at 90\% confidence, which corresponds to 3.8 kpc at a redshift of 0.1. This resolution is similar to the half-light radius of a typical galaxy \citep[e.g.,][]{2003MNRAS.343..978S}. Thus, in many cases, DSA-110 FRBs can be associated with individual galaxies at the characteristic distance of the FRB population. However, by requiring robust association with PS1, we introduce an optical magnitude limit to the FRB host sample. Selection requires that the host be detected in r and i bands above the $5\sigma$\ detection limits of 23.2 and 23.1 mag \newtext{, respectively}.

\subsection{Associating FRBs to Host Galaxies}
\label{sec:assoc}

Table \ref{tab:hostbasic} summarizes the properties of the host galaxies used to associate them with the FRBs. We use \texttt{astropath} \citep{2021ApJ...911...95A} to calculate the association probability for each FRB to its host galaxy. The Bayesian framework estimates the association probability for all nearby galaxies from the FRB position and error, as well as galaxy magnitude, location, and size. Galaxies with an association probability greater than 0.9 are considered robust, given the conservative assumptions, and thus included in this sample.

We use the adopted priors in \citet{2021ApJ...911...95A}, defined as an exponential FRB angular offset distribution and an association probability that scales inversely to the number density of galaxies at a given magnitude (``exp'' and ``inverse'', respectively). We deviate from the adopted set of \texttt{astropath} priors by using a non-zero prior on undetected host galaxies, $P(U)$. \citet{2021arXiv211207639S} analyzed the false-positive and false-negative association rate for simulated FRB host galaxies. At the depth of PS1 $r$-band imaging, a reliable and accurate association probability is found for $P(U)=0.2$. All 11 FRBs are covered by Pan-STARRS1 (DR2), but six of the FRBs are covered by the deeper Legacy survey \citep[DR9;][]{2019AJ....157..168D}. For FRBs covered by the Legacy survey, we use $P(U)=0.1$.

For each FRB, we build a galaxy catalog from PS1 stack images or Legacy catalogs by selecting for resolved sources within 30\arcsec \newtext{(117 kpc at the median redshift of this sample; see Table \ref{tab:hostderived})}. For PS1, point-like sources are removed if they are in the PS1-PSC catalog with ``ps\_score'' less than 0.83 \citep{2018PASP..130l8001T}. No compactness filter was used for Legacy Survey sources, but all FRB host galaxies are resolved, with $R_e>1\arcsec$. For PS1, we use the r-band Kron radius to represent the galaxy's size, which is within 20\% of the half-light radius under realistic scenarios \citep{2005PASA...22..118G}. \newtext{We find an association probability much greater than 0.9 for all FRBs presented in Table \ref{tab:hostbasic}.}

\begin{table*}
\begin{tabular}{l|ccccccc}
FRB Name & Host Name & Host RA & Host Dec & m$_{\rm{r}}$ & r$_{\rm{e}}$ & Source & P$_{\rm{host}}$ \\
 & & (J2000) & (J2000) & (mag) & (arcsec) & & \\ \hline
\frbzach & PSO J310.1977$+$72.8826 & 20:40:47.419 & +72:52:57.90 & \newtext{16.15 $\pm$ 0.01} & 8.356 & P & 0.99 \\
\frbalex & PSO J350.8747$+$72.1918 & 23:23:29.925 & +72:11:30.80 & \newtext{20.45 $\pm$ 0.05} & 1.393 & P & 0.99 \\
\frbwhitney & PSO J134.7211$+$73.4910 & 08:58:53.054 & +73:29:27.45 & \newtext{21.20 $\pm$ 0.08} & 1.909 & P & 0.99 \\
\frbmark & IRAS 02044$+$7048 & 02:08:40.812 & +71:02:09.65 & \newtext{13.25 $\pm$ 0.01} & 11.764 & P & 0.99 \\
\frbquincy & PSO J219.1065$+$70.0953 & 14:36:25.584 & +70:05:43.60 & \newtext{22.22 $\pm$ 0.20} & 1.290 & L & 0.97 \\
\frboran & PSO J318.0447$+$72.8272 & 21:12:10.727 & +72:49:37.76 & \newtext{19.91 $\pm$ 0.16} & 2.425 & P & 0.98 \\
\frbjackie & PSO J282.6748$+$70.2427 & 18:50:41.916 & +70:14:33.95 & \newtext{16.51 $\pm$ 0.01} & 3.938 & L & 0.99 \\
\frbansel & PSO J311.9820$+$72.5848 & 20:47:55.6 & +72:35:06.5 & \newtext{19.36 $\pm$ 0.16} & 1.935 & P & 1.0 \\
\frbelektra & PSO J282.0585$+$73.3363 & 18:48:13.958 & +73:20:10.70 & \newtext{20.05 $\pm$ 0.04} & 1.278 & L & 0.97 \\
\frbetienne & PSO J240.2568$+$70.9180 & 16:01:01.634 & +70:55:05.05 & \newtext{19.46 $\pm$ 0.03} & 2.116 & L & 0.99 \\
\frbjuan & PSO J280.8014$+$70.5242 & 18:43:12.3 & +70:31:27.2 & \newtext{19.67 $\pm$ 0.03} & 3.096 & L & 1.0 \\
\end{tabular}
\caption{Parameters of DSA-110 FRB Host galaxies, listed in discovery order. The columns include information used to identify and associate the galaxy to the FRB, including r-band magnitude and effective radius\newtext{, r$_{\rm{e}}$}. $P_{\rm{host}}$ is the \texttt{astropath} posterior probability of association of the galaxy to the FRB. The field ``source'' describes what catalog was used to search for potential counterparts, with ``P'' for Pan-STARRS1 and ``L'' for Legacy Survey. P$_{\rm{host}}$ refers to the probability that the FRB is associated to this galaxy (see \S \ref{sec:assoc}).}
\label{tab:hostbasic}
\end{table*}

\subsection{Optical/IR Photometry}

Table \ref{tab:details} summarizes the photometric and spectroscopic measurements made and used in modeling of host galaxies. We \newtext{use catalog photometry from}\oldtext{performed photometry on images from} wide-area OIR surveys PS1, Two Micron All Sky Survey~\citep[2MASS;][]{2006AJ....131.1163S} and ALLWISE~\citep{2014yCat.2328....0C} surveys. We also acquired IR photometric data in $J$ and $H$ bands with the Wide Field Infrared Camera \citep[WIRC;][]{2003SPIE.4841..451W} observing instrument mounted on the Palomar 200-inch Telescope. These observations were acquired on August 16, 2022, and were reduced by a custom pipeline following the standard data reduction pipeline procedures. We largely follow the analysis framework established in \citet{2023arXiv230214782S} for characterizing the host galaxies of these FRBs (including removal of a star on the host galaxy of \frbjackie). We \oldtext{execute isophotal analysis on these host galaxies in PS1 i-band images, where the center and the size of the galaxy are left as free parameters. We identify the best isophote as the one that captures $\gtrsim$95\% of the galaxy. We further scale this best-fit isophote by the point spread function of different photometric bands for performing aperture photometry.} \newtext{convert the Vega magnitudes to AB magnitudes and correct for galactic dust extinction.}



\subsection{Optical/IR Spectroscopy}

\newtext{With a reliable host identified, we observed each galaxy with optical/IR spectroscopic instruments. Detection of line emission allows} \oldtext{Optical spectroscopy is crucial for} accurate redshift measurement, line-flux estimates to better understand the nature of ongoing star-formation and nuclear activity, and for modeling the ages of the stellar populations in a galaxy. The age of the stellar population of an FRB host galaxy \newtext{tells us about how star formation creates FRB sources, a crucial test of origin models.}\oldtext{, and preferably host environment, encodes the age of the probable progenitor object, which can be determined by probing host-galaxy star-formation histories.}
Table \ref{tab:details} summarizes the spectroscopic instruments used for this FRB sample.

We observed with the Low-Resolution Imaging Spectrometer on the Keck I telescope~\citep[Keck-I/LRIS][]{1995PASP..107..375O} on 26 May, 21 July, and 17 October of 2022. \oldtext{On the last of these observing sessions, the red component of the detector had malfunctioned. We used a mirror to direct light into the blue arm, and} The light was dispersed using a 300/5000 grism to measure a detector-frame wavelength range from 3450 to 6700\AA. The spectra were reduced with the standard \sw{lpipe} software~\citep{2019PASP..131h4503P} and calibrated using observations of the standard star BD+28 4211. The spectrum was scaled to match the PS1 photometry to account for slit losses. 

We observed with the Double Spectrograph~\citep[DBSP;][]{1982PASP...94..586O} mounted at the Palomar Hale 200-inch telescope on 1 June, 12 December of 2022 and 28 May 2023. Data were reduced following procedures described in \citet{2022MNRAS.513..982R}. 

We measure the spectroscopic redshift and the line fluxes of the host galaxies using the penalized PiXel-Fitting software~\citep[\sw{pPXF};][]{2017MNRAS.466..798C, 2022arXiv220814974C} by jointly fitting the stellar continuum and nebular emission using the MILES stellar library~\citep{2006MNRAS.371..703S}. \newtext{The line fluxes are shown in Table \ref{tab:lines}. A redshift could be reliably identified for all hosts by the detection of at least 3 emission lines. For one FRB, \frbjackie, we measured the redshift by modeling absorption lines.}

\subsection{Radio and High-Energy Counterpart Search}

\newtext{Two FRBs have been associated with PRS and another associated with a known high-energy source (a magnetar). We searched for new examples of these associations by matching FRB positions to wide-area radio and X-ray catalogs. The focus was on compact counterparts, which are simpler to characterize and associate with point-like FRB emission.}\oldtext{We searched for compact radio and X-ray counterparts in catalogs and images associated with large surveys.} Radio catalogs with coverage of this declination range include VLASS \citep{2020PASP..132c5001L}, NVSS \citep{1998AJ....115.1693C}, TGSS \citep{2017A&A...598A..78I}, GB6 \citep{1996ApJS..103..427G}, WeNSS \citep{1997A&AS..124..259R}, VLSSr \citep{2014MNRAS.440..327L}, and LoTSS \citep{2019A&A...622A...1S}. At high energies, we searched the Chandra Source Catalog 2 \citep[``csc21\_snapshot\_scs'';][]{2020AAS...23515405E} and XMM/Newton 4XMM-DR11 \citep{2020A&A...641A.136W}. No counterpart was found for any of our sample in these catalogs. \newtext{Radio limits are fairly heterogeneous and range from approximately 1 mJy to tens of mJy over frequencies from 120 MHz to 5 GHz. X-ray limits. Most FRBs had no coverage in high-energy catalogs.}

As the nearest non-repeating FRB, the environment and counterparts to \frbmark\ can be probed especially well. We observed \frbmark\ with the Karl G. Jansky Very Large Array (VLA) under program 22A-490 to search for \newtext{PRS}\oldtext{a persistent radio counterpart} and repeat bursts. We observed with antennas in the A configuration in the 1.4 and 5 GHz bands. The correlator was configured for continuum observing and a commensal 10-ms transient search with the \textit{realfast} instrument \citep{2018ApJS..236....8L}. We detected no burst or persistent counterpart and the continuum imaging sensitivity was approximately 21 microJy/beam in both bands (1$\sigma$, dual-polarization, robust weighting).

\section{Analysis}
\label{sec:ana}

\subsection{Burst Properties}
\label{sec:bursts}

\begin{figure}
    \centering
    \includegraphics[width=0.45\textwidth]{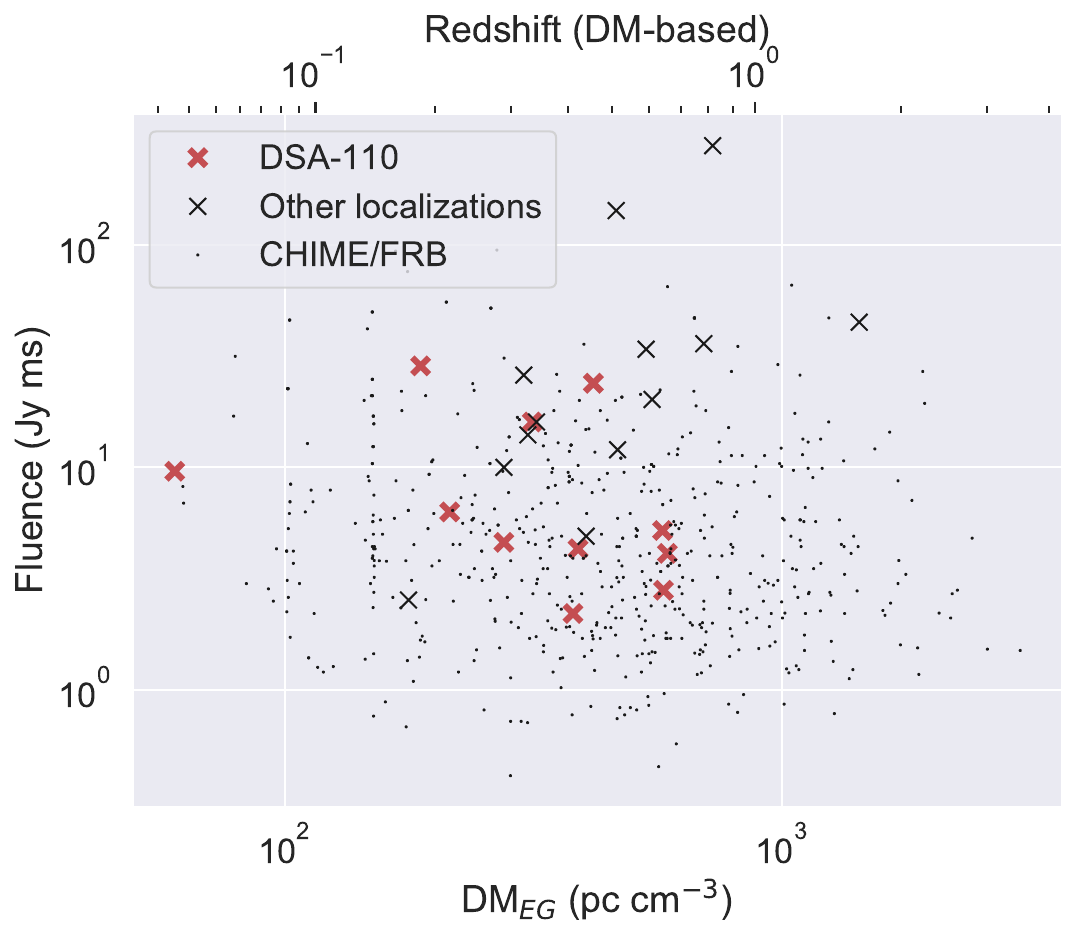}
    \includegraphics[width=0.45\textwidth]{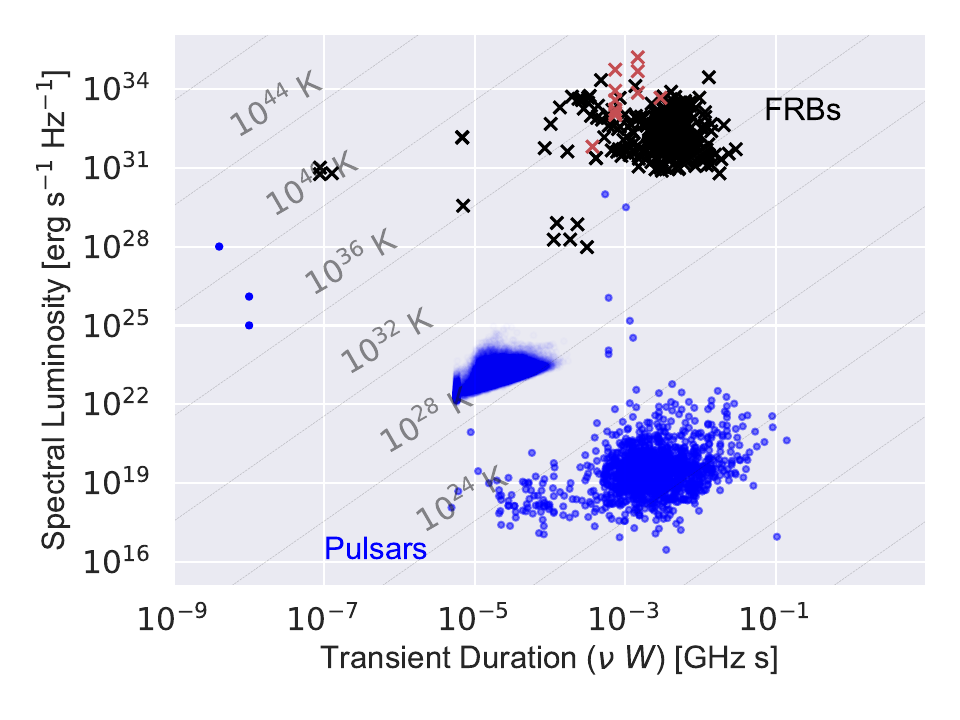}
    \caption{(Left) Burst DM versus fluence for a sample of DSA-110 (red crosses), other localized events (black crosses) and CHIME-FRB detections \citep[black points;][]{2021ApJS..257...59C}. The scale of DM to redshift assumes linear convertion with a Milky Way halo contribution of 30 pc cm$^{-3}$\ and 850 pc cm$^{-3}$ to redshift of 1. (Right) Luminosity versus timescale for Galactic (blue circles) and extragalactic (black crosses) fast transients. The DSA-110 FRB sample is shown in red crosses.}
    \label{fig:burstprops}
\end{figure}

The radio bursts themselves can be used to constrain both intrinsic properties and propagation effects. Bursts are characterized by DM, intrinsic width and structure, scintillation, scattering, as well as polarimetric properties. For the total-intensity burst properties considered here, we find that the DSA-110 sample is typical of the non-repeating population identified previously. Figure \ref{fig:burstprops} shows the extragalactic DM compared to burst fluence for DSA-110 and other FRB discoveries. The new sample has a median fluence of $\sim$5 Jy ms, compared to $\sim$20 Jy ms for the reference sample of localized FRBs with data available in \citet{10.5281/zenodo.8125230}. \newtext{This reflects the improved sensitivity of DSA-110 relative to other localization instruments (dominated by ASKAP).}

Figure \ref{fig:burstprops} also shows the luminosity and timescale distribution for Galactic and extragalactic millisecond transients \citep[c.f.][]{2022NatAs...6..393N}. The present sample is somewhat shorter and more luminous than the bulk of previous detections. This offset reflects the sensitivity and relatively fast sampling time of the DSA-110 search system. We note that the requirement for a PS1 host biases against very distant hosts and high FRB luminosities. Future FRB samples from the DSA-110 with dedicated follow-up optical observing should be expected to show higher burst luminosities.

Of special interest is whether bursts repeat or have complex ``substructure'' within individual bursts. 
Using the burst classification defined in \citet{2021ApJ...923....1P}, we characterize 8 bursts as ``broad'' and three as ``complex''. These terms are synonymous with having single or multiple millisecond-scale components in time, respectively. We note that the relatively fast time resolution and relatively narrow bandwidth of DSA-110 makes it easier for DSA-110 to detect and identify bursts as complex, while limiting the ability to see narrow or downward drifting bursts. A more complete discussion of burst classification with the inclusion of polarimetric properties is presented in \newtext{\citet{2023arXiv230806813S}}\oldtext{Sherman et al. (in prep)}, and a detailed analysis of the total-intensity properties will be presented in Chen et al. (in prep).

None of the bursts were observed to repeat on timescales longer than a second. Based on the observing pattern during the science commissioning period, the FRB locations were nominally observed for 150 hours through the 2022 observing year\footnote{The effective observing time is currently not well constrained, so no rate estimate can be calculated.}. If the three multi-component bursts are treated as repeat bursts, then the wait times range from 0.5 to 3 ms. This is consistent with first peak of the wait time distribution for active repeating FRBs such as 121102 \citep{10.1038/s41586-021-03878-5}.


\subsection{Host Galaxy SED Modeling}
\label{sec:sed}

We obtain the stellar properties of our set of host galaxies using the stellar population synthesis modeling software \sw{Prospector}~\citep{2021ApJS..254...22J}. We follow the SED modeling approach described in \citet{2023arXiv230214782S}, which includes a discussion of spectral processing, model parameters, and priors. \newtext{In brief, we simultaneously fit a model to the spectro-photometric data. The redshift is initialized to the value measured with \sw{pPXF} and given a uniform prior width of 1\%. We use a continuity non-parametric star formation history with 7 bins and the Chabrier initial mass function. We add an AGN component to the model whenever data above 2$~\mu$m is available. We sample from the posterior using the ensemble sampler \sw{emcee}~\citep{2013PASP..125..306F}.}

Generally, the photometric and flux-calibrated spectroscopic observations are jointly fit when possible. \oldtext{The host galaxies of \frbzach, \frboran, and \frbmark\ have low signal-to-noise ratios in their spectra and thus the absorption features are not well characterized, due to which we refrain from doing a joint photometric-spectroscopic fit. Therefore, for these three hosts, we only do a photometric SED modeling with similar procedures to \citet{2023arXiv230214782S}, but drop out spectroscopic calibrations from the model parameters.} \newtext{Spectroscopy toward host galaxies of \frbzach, \frboran, and \frbmark\ had poor quality, so channels were selected around line emission for inclusion in SED modeling.} For \frbquincy, we use legacy survey data instead of PS1 in our SED modeling due to better SNR in legacy data. The resulting SED fits for all the host galaxies, along with their constrained non-parametric star formation histories, are displayed in Figure~\ref{fig:SED_fits} and Figure~\ref{fig:SFH_fits}, respectively. The set of derived host properties is summarized in Table~\ref{tab:hostderived} and plotted in Figure \ref{fig:hosts2d}. Best-fit line fluxes \newtext{measured by pPXF} are shown in Table \ref{tab:lines}. We note that the recent star formation history is not well constrained in a few cases. We suspect that this is primarily due to the lack of UV photometry in our SED modeling.

\begin{table*}
\begin{tabular}{l|ccccccc}
FRB name & Redshift & M$_{\rm{r}}$ & g-r & log(M$_*$) & SFR$_{\rm{100 Myr}}$ & t$_{\rm{m}}$ & L$_{\rm{PRS}}$ \\
       &               & (mag) & (mag) & (Msun) & (Msun/yr) & (Gyr) & (erg/s/Hz) \\ \hline
       
\frbzach & 0.043040    & \newtext{$-20.20_{-0.00}^{+0.00}$} & \newtext{$0.30_{-0.00}^{+0.01}$}  & \newtext{$9.91_{-0.02}^{+0.03}$} & \newtext{$1.16_{-0.08}^{+0.10}$} & \newtext{$6.53_{-0.72}^{+0.84}$} & $<2.3\times10^{28}$ \\

\frbalex & 0.248123    & \newtext{$-19.97_{-0.00}^{+0.00}$} & \newtext{$0.85_{-0.01}^{+0.01}$} & \newtext{$10.04_{-0.01}^{+0.02}$} & \newtext{$2.71_{-0.61}^{+0.76}$} & \newtext{$0.65_{-0.04}^{+0.04}$} & $<1.0\times10^{30}$ \\

\frbwhitney & 0.477958 & \newtext{$-20.79_{-0.03}^{+0.03}$} & \newtext{$0.99_{-0.05}^{+0.06}$}  & \newtext{$9.97_{-0.05}^{+0.05}$} & \newtext{$4.25_{-1.67}^{+2.45}$} & \newtext{$2.13_{-0.60}^{+1.06}$} & $<4.6\times10^{30}$ \\


\frbmark & 0.011228    & \newtext{$-20.23_{-0.01}^{+0.01}$} & \newtext{$0.46_{-0.02}^{+0.03}$} & \newtext{$10.16_{-0.06}^{+0.03}$} & \newtext{$0.21_{-0.05}^{+0.12}$} & \newtext{$6.30_{-1.82}^{+1.51}$} & $<1.8\times10^{26}$\tablenotemark{a} \\

\frbquincy & 0.622000  & \newtext{$-20.88_{-0.01}^{+0.01}$} & \newtext{$0.64_{-0.03}^{+0.04}$} & \newtext{$10.31_{-0.05}^{+0.06}$} & \newtext{$29.41_{-10.70}^{+18.28}$} & \newtext{$4.62_{-0.98}^{+1.20}$} & $<8.6\times10^{30}$ \\

\frboran & 0.30039    & \newtext{$-20.63_{-0.01}^{+0.01}$}	& \newtext{$1.50_{-0.03}^{+0.03}$} & \newtext{$10.29_{-0.07}^{+0.06}$} & \newtext{$5.08_{-0.47}^{+0.31}$} & \newtext{$0.07_{-0.01}^{+0.02}$} & $<1.5\times10^{30}$ \\

\frbjackie & 0.089400  & \newtext{$-21.34_{-0.00}^{+0.00}$} & \newtext{$0.78_{-0.00}^{+0.00}$} & \newtext{$10.79_{-0.01}^{+0.01}$} & \newtext{$0.23_{-0.03}^{+0.04}$} & \newtext{$9.03_{-0.42}^{+0.46}$} & $<1.1\times10^{29}$ \\

\frbansel & 0.241397   & \newtext{$-19.46_{-0.02}^{+0.07}$}	& \newtext{$0.87_{-0.16}^{+0.06}$} & \newtext{$9.95_{-0.15}^{+0.17}$} & \newtext{$1.62_{-0.31}^{+0.47}$} & \newtext{$3.46_{-1.02}^{+1.49}$} & $<9.4\times10^{29}$ \\

\frbelektra & 0.113900 & \newtext{$-18.31_{-0.02}^{+0.02}$} & \newtext{$0.52_{-0.02}^{+0.03}$} & \newtext{$9.48_{-0.09}^{+0.07}$} & \newtext{$0.72_{-0.15}^{+0.20}$} & \newtext{$7.50_{-1.50}^{+1.86}$} & $<1.8\times10^{29}$ \\

\frbetienne & 0.158239 & \newtext{$-19.96_{-0.01}^{+0.02}$} & \newtext{$0.48_{-0.01}^{+0.02}$} & \newtext{$9.81_{-0.04}^{+0.08}$} & \newtext{$4.10_{-1.00}^{+1.09}$} & \newtext{$3.86_{-0.89}^{+1.56}$}  & $<3.6\times10^{29}$ \\

\frbjuan & 0.284669    & \newtext{$-20.94_{-0.01}^{+0.01}$} & \newtext{$1.50_{-0.01}^{+0.01}$} & \newtext{$10.99_{-0.02}^{+0.02}$} & \newtext{$0.15_{-0.03}^{+0.04}$} & \newtext{$6.75_{-0.41}^{+0.56}$} & $<1.4\times10^{30}$ \\    
\end{tabular}
\caption{Measured and derived properties of the DSA-110 FRB host galaxies with FRBs listed in discovery order. $t_{\rm{m}}$ refers to the mass-weighted age of the galaxy stellar population. See Section \ref{sec:spres} for model fit plots. Limit on persistent radio counterparts derived from the VLA Sky Survey at 3 GHz, except where noted.}
\tablenotetext{a}{Dedicated VLA follow up at 1.4 and 5 GHz.}
\label{tab:hostderived}
\end{table*}

\subsection{Limits on Persistent Radio Sources}

PRS \oldtext{Persistent radio sources are luminous compact radio sources that are spatially coincident with FRBs. They}
have been robustly associated with two FRBs \citep{2017ApJ...834L...7T,2022Natur.606..873N} and have been a valuable driver of models for the physics of FRBs \citep[e.g.,][]{2018ApJ...868L...4M}. \newtext{\citet{2022ApJ...927...55L} presents a phenomenological definition of a PRS as a compact (sub-parsec), luminous (L$_r>10^{29}$\ erg s$^{-1}$\ Hz$^{^-1}$) radio source that is spatially coincident with an FRB. Given this definition, they were able to provide a more homogenous analysis of the PRS population and how they may be causally connected to FRBs. Note that this luminosity limit is defined phenomenologically to exclude astrophysical foregrounds and that non-detections do not necessarily rule out the physical object that produces PRS-like emission.}

We find no persistent radio counterparts associated with these DSA-110 FRBs. The VLASS $3\sigma$~ limit is 0.5 mJy (at 3 GHz). Using the PRS luminosity threshold proposed previously \oldtext{\citep[L$_r>10^{29}$\ erg s$^{-1}$\ Hz$^{^-1}$;][]{2022ApJ...927...55L}}, we can exclude the presence of a PRS in FRBs \frbzach\ and \frbmark. FRBs \frbjackie, \frboran, and \frbelektra\ are slightly above the limit, but still fainter than the two confirmed PRS. \oldtext{Note that this luminosity limit is mostly defined to exclude astrophysical foregrounds and that non-detections do not necessarily rule out PRS-like emission.}

We reviewed recently published FRB localizations to build on the sample of known PRS and upper limits \citep{2023arXiv230205465G,2023arXiv230209787D,2022arXiv221004680R,2023ApJ...949L...3R,2023arXiv230209754C}. Among new localizations from ASKAP and MeerKAT, we find that non-repeating FRBs 20211212A, 20211127I, 20210405I, and 20210410D all have significant constraints on PRS counterparts. FRBs presented in \citet{2022MNRAS.514.1961R} have less robust associations to host galaxies and \citet{2022MNRAS.515.1365C} includes a PRS candidate that may be attributed to \newtext{star formation in} its host galaxy, so they are not counted here.

Considering these new \newtext{PRS constraints from DSA-110, MeerKAT, and ASKAP}\oldtext{measurements}, we find 15 non-repeating FRBs with upper limits on PRS counterparts. The measurements toward repeating FRBs remain unchanged, with 2 detections out of 6 useful measurements. Repeating the analysis of \citet{2022ApJ...927...55L}, we \oldtext{place a} \newtext{lower the} 90\% upper limit on the fraction of non-repeating FRBs with a PRS at 0.14, \newtext{and find an 8\% chance that the repeater and non-repeater PRS fraction are consistent with each other.}\oldtext{while the bounds on the repeater fraction are unchanged. The chance that the subset of repeating FRBs could randomly be associated with the two PRS sources is 8\%.} We conclude that there is weak evidence for repeating FRBs to be associated with PRS. \newtext{We encourage more radio imaging of localized FRBs to understand whether PRS and repetition have a causal relationship}.


\section{Discussion}
\label{sec:dis}

\subsection{What Kind of Galaxies Host FRBs?}

Analysis of \oldtext{the host galaxies of} a population of \oldtext{extragalactic sources}\newtext{FRB host galaxies} can provide insights into formation and evolution channels. We use the present DSA-110 sample of localized FRBs to first explore the characteristics of the host galaxies. In \S~\ref{sec:dtd} below, we focus on the star-formation histories as probes of the formation of FRB progenitors via the delay time distribution.  

\begin{figure}
    \centering
    \includegraphics[width=0.49\textwidth]{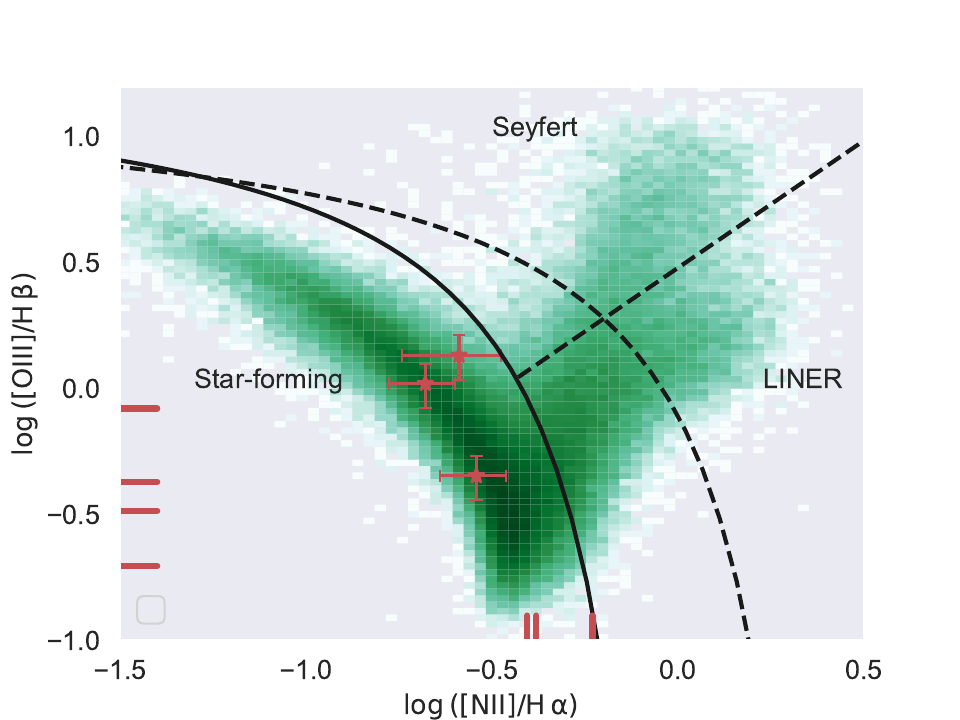}
    \includegraphics[width=0.49\textwidth]{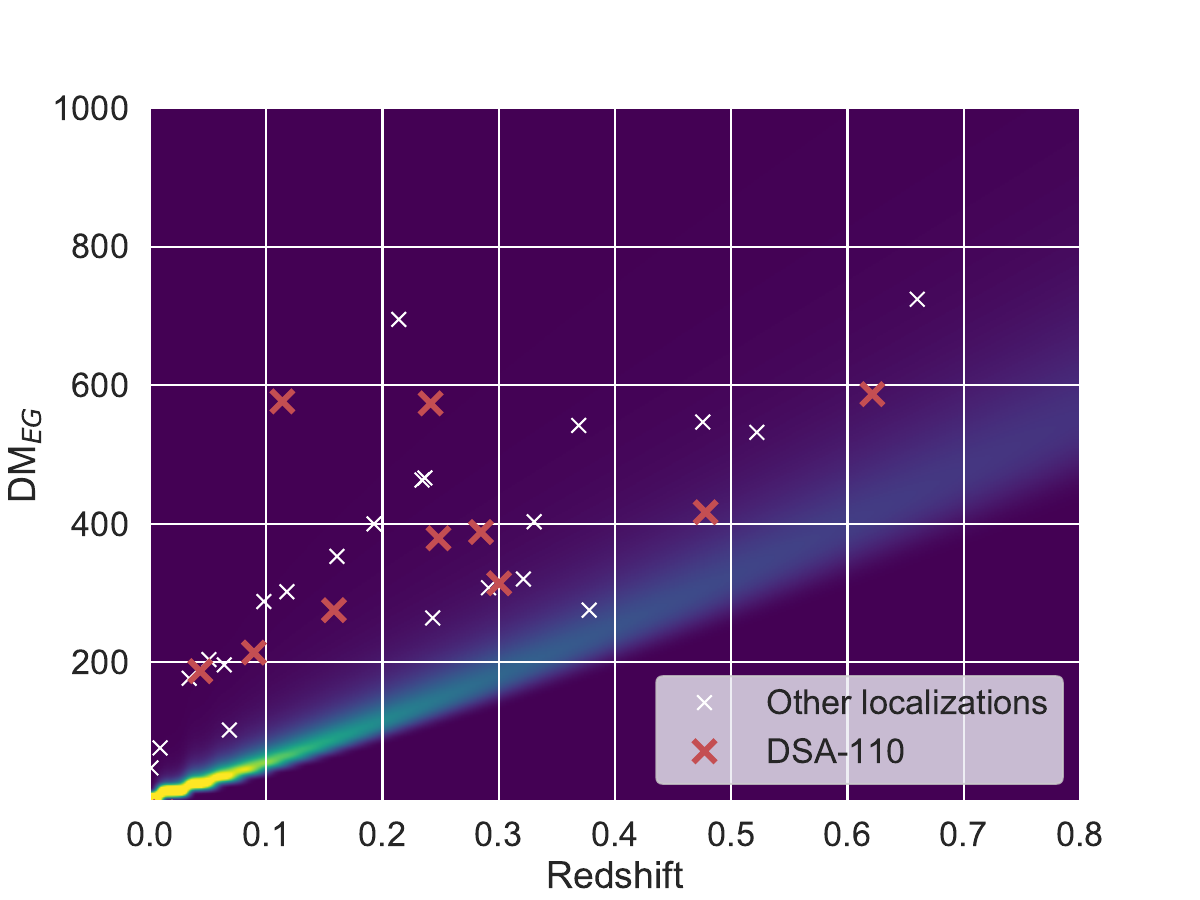}
    \caption{(Left) BPT diagram of nebular emission from DSA-110 host galaxies \newtext{plotted over line ratios measured for field galaxies in SDSS}. \newtext{The solid line shows the boundary on star-formation ionization \citep{2003MNRAS.346.1055K}, the linear dashed line shows the LINER/Seyfert boundary \citep{2010MNRAS.403.1036C}, and the dashed curved line shows the theoretical star-formation boundary defined in \citet{2001ApJ...556..121K}.} Three DSA-110 host galaxies have all four transitions measured and seven have a single line ratio measured. The latter are shown as red lines at the edge of the plot. \frbjackie\ has no detected emission lines and is not shown.
    (Right) Extragalactic DM for FRBs compared to the probability density of DM from the intergalactic medium versus redshift. DSA-110 FRB sample shown with cyan stars and published events with white crosses. We use the FRB code with function \texttt{prob\_dmz.grid\_P\_DMcosmic\_z}\ with \texttt{beta=3.0}\ and \texttt{F=0.31}. }
    \label{fig:hostprops}
\end{figure}

The basic properties of the host galaxies of the DSA-110 sample, and the FRB sightlines through these galaxies, are not unusual with respect to the existing sample of localized FRBs \citep[e.g,][]{2023arXiv230205465G}. The left panel of Figure \ref{fig:hostprops} shows the flux ratios of nebular emission lines observed from the DSA-110 hosts, in a BPT diagram \citep{1981PASP...93....5B}. Most measurements and constraints on the line ratios are consistent with nebular emission driven by star-formation activity, and we see no compelling evidence for AGN contributions. We note that \frbmark\ is more consistent with line ratios seen in LINER galaxies, although that line ratio was measured directly on the nucleus of the (well resolved) galaxy \citep{2023arXiv230101000R}. \newtext{Despite line ratios consistent with ionization by star-formation, we found that SED modeling was improved by the inclusion of AGN terms; IR colors are particularly sensitive to the presence of AGN \citep{2018ApJ...854...62L}.}\oldtext{This BPT analysis justifies the standard techniques used for stellar population synthesis modeling.}

The right panel of Figure \ref{fig:hostprops} compares the published and present sample of FRBs to a model of DM from the intergalactic medium. The median redshift of the DSA-110 sample is 0.24. \oldtext{This}\newtext{A typical} analysis assumes $\rm{DM} = \rm{DM}_{\rm{ISM,MW}} + \rm{DM}_{\rm{halo,MW}} + \rm{DM}_{\rm{IGM}} + \rm{DM}_{\rm{host}}/(1+z)$. This \oldtext{model is a reasonable representation of}\newtext{equation explains} the DM-redshift relationship (``Macquart relation'') \newtext{when $\rm{DM}_{\rm{IGM}}$ is evaluated based on the mean baryon density over the path length to the FRB redshift, $z$.}\oldtext{, and no egregious outliers are present in the DSA-110 sample relative to published FRBs, although} There are four DSA-110 FRBs with \oldtext{some} more than 200 pc cm$^{-3}$\ excess DM (\frbalex, \frbansel, \frbelektra, \frbjuan). In the case of \frbelektra, this is directly attributed to the intracluster gas associated with the host galaxy \citep{arXiv:2302.14788}\oldtext{, and} Contributions from intervening structures (e.g., filaments, groups) \oldtext{are not fully characterized} for other DSA-110 FRBs\newtext{will be discussed in more detail in Connor et al (in prep)}.

Useful morphological information is only accessible for a handful of DSA-110 hosts. As discussed by \citet{2023arXiv230101000R}, the host of \frbmark\ is a face-on barred spiral and the FRB is significantly offset. The disk-dominated host galaxy of \frbzach\ is viewed nearly edge-on, and a significant $\rm{DM}_{\rm{IGM}} + \rm{DM}_{\rm{host}}/(1+z)=170$\,pc\,cm$^{-3}$ is observed (assuming $\rm{DM}_{\rm{halo,MW}}=10$\,pc\,cm$^{-3}$) despite the low redshift of 0.043. Detailed morphological analysis will be presented in Sharma et al. (in prep).

\begin{figure}
    \centering
    \includegraphics[width=0.345\textwidth]{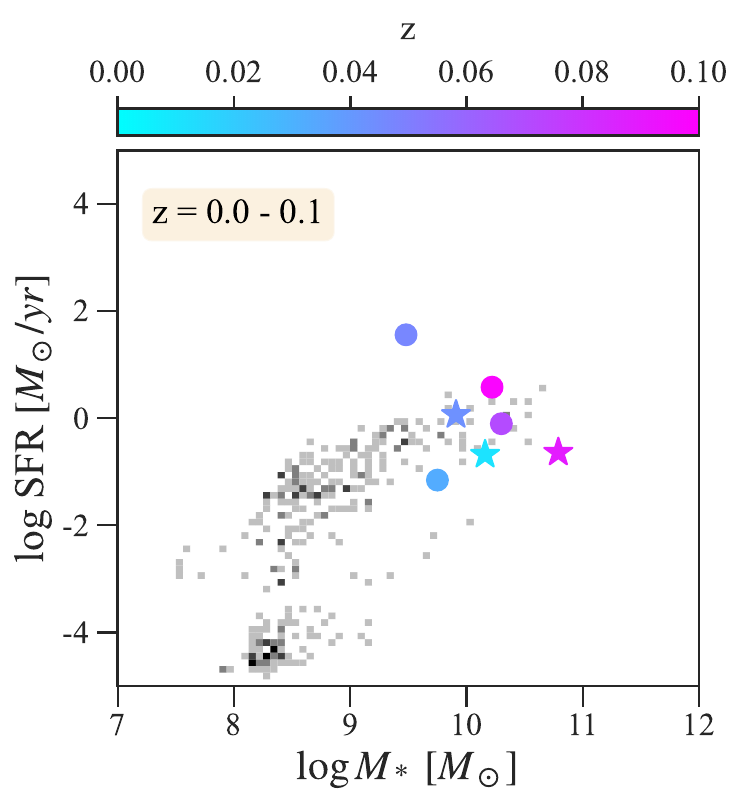}
    \includegraphics[width=0.31\textwidth]{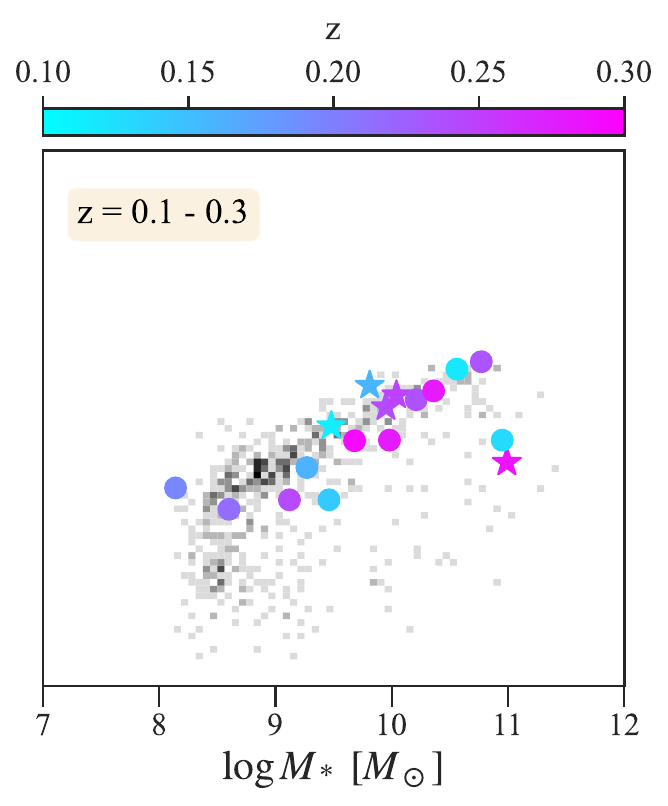}
    \includegraphics[width=0.302\textwidth]{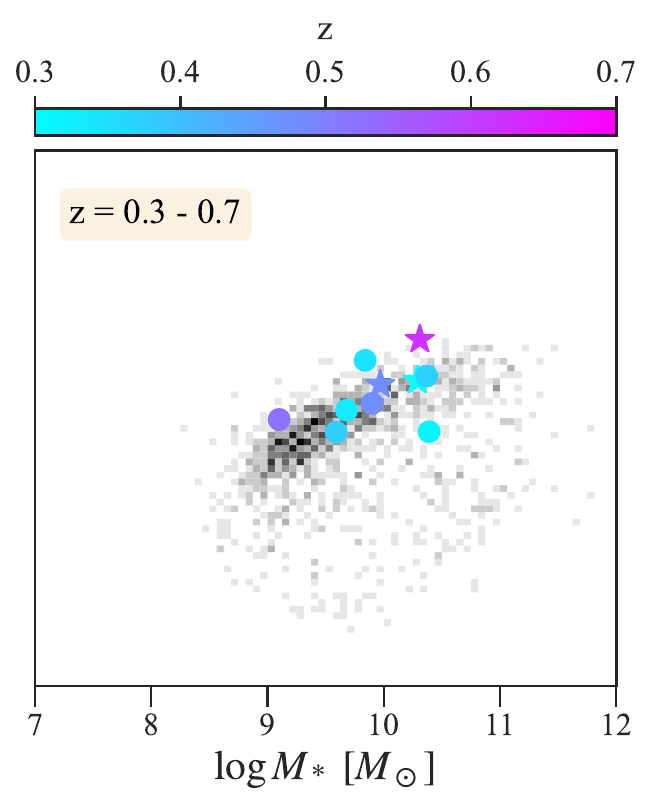}
    \caption{Stellar mass versus recent star-formation rate for FRB host galaxies and COSMOS galaxies in three redshift bins, $0-0.1$, $0.1-0.3$, and $0.3-0.7$ \citep{2016ApJS..224...24L}. The background color shows the number density of galaxies, which is dominated by the star-forming sequence. The FRB hosts are marked by circles were presented in \citet{2023arXiv230205465G}, and hosts marked by stars are discovered by DSA-110 and presented here. The colormap for the symbols shows the redshift of the FRB host galaxy.}
    \label{fig:hosts2d}
\end{figure}

Important insight can be gained from an analysis of the total stellar mass and recent star-formation rates of the FRB host-galaxy sample. Figure \ref{fig:hosts2d} shows the stellar mass and star-formation rate for both the published and DSA-110 FRB host galaxies in three redshift bins. The star-formation rates are averaged over the last 100 Myr from a non-parametric star-formation history analysis. For reference, we used the largest field galaxy sample modeled using similar techniques \citet[COSMOS;][]{2020ApJ...893..111L}, following \citet{2023arXiv230205465G}, and the published FRB hosts were also analyzed using similar techniques. The DSA-110 sample spans two orders of magnitude in both stellar mass and star-formation rate, and three orders of magnitude in specific star-formation rate. This diversity of host properties is consistent with the published FRB sample.   

The DSA-110 sample is subject to a selection effect on optical magnitude. For example, using \texttt{kcorrect} \citep{2007AJ....133..734B}, we estimate the characteristic galaxy that is detectable in PS1 ($5\sigma$ stack limits). For galaxies with mass-to-light ratios from 0.7 to 1, we find minimum stellar masses, $M_{\odot, \rm{min}}$ of $8\times10^8 M_{\odot}$ at $z=0.1$, $9\times10^9 M_{\odot}$ at $z=0.3$, and $4\times10^{10} M_{\odot}$ at $z=0.6$. This selection effect is clear in Figure \ref{fig:hosts2d}. While the minimum stellar mass of a DSA-110 host excludes dwarf galaxies, the present sample is sensitive to the star-forming main sequence and quiescent galaxies in all redshift bins.

\newtext{Figure~\ref{fig:hosts2d} shows the distribution of mass and SFR for published and DSA-110 host galaxies.} Early analysis of FRB hosts \citep{2020ApJ...903..152H, 2022AJ....163...69B} seemed to \oldtext{identify FRB hosts as offset from the star-forming main sequence and more closely associated with the ``green valley'' galaxies with slowing star-formation} \newtext{distinguish FRB hosts from the star-forming main sequence and quiescent galaxy population} \citep{2014MNRAS.440..889S}. However, newer FRB host modeling analysis that \oldtext{compares against}\newtext{uses} non-parametric star formation histories shows no such \oldtext{offset}\newtext{distinction} \citep{2019ApJ...877..140L,2023arXiv230205465G}. Figure \ref{fig:hosts2d} confirms that FRB host galaxies \oldtext{are not associated with green valley galaxies. Instead, we associate}\newtext{are associated with} star-forming and quiescent hosts, which dominate the field galaxy population by number and stellar mass.

\newtext{A}\oldtext{The major} new result from the DSA-110 FRB host-galaxy sample is the \oldtext{prevalence}\newtext{identification} of massive hosts with low specific star-formation rates \citep[as parameterized in, e.g.,][]{2014ApJS..214...15S}. \oldtext{Four}\newtext{Two} of 11 hosts in our sample \oldtext{have stellar masses greater than 10$^{11}$\ M$_{\odot}$, and all have specific star-formation rates below the star-forming main sequence.} \newtext{(\frbjackie, \frbjuan) are massive, low SFR galaxies with high mass-weighted stellar ages. However, the significance of the quiescent hosts is only meaningful when considered relative to the incidence of star-forming hosts. The fact that the DSA-110 hosts are a magnitude-limited sample means that we can test this significance statistically.}\oldtext{This preference for massive hosts initially seems to contrast with the published sample of FRB hosts . Although our sample is defined by detection in PS1 and association using priors that favor brighter galaxies, our method does not deviate from that used in the community and is unlikely to be biased for the stellar masses under consideration.  The DSA-110 is a more sensitive FRB detector than other instruments capable of interferometric localization, and our FRB sample is fainter than other localizations (see Figure). However, for this to be a source of bias would require apparently fainter FRBs to be associated with more massive hosts, something that we do not consider likely pending further investigation. Selection effects, particularly due to radio propagation, may lead to a preference for lower specific star-formation rates, but that bias is not unique to the present sample. Considering that there is no known bias and that the number of massive host galaxies is still small, we conclude that the DSA-110 sample is the first that is large and uniform enough to reveal these rarest kinds of host. We proceed by considering the implications of the massive hosts among the DSA-110 sample alone.}

\begin{figure}
    \centering
    \includegraphics[width=\textwidth]{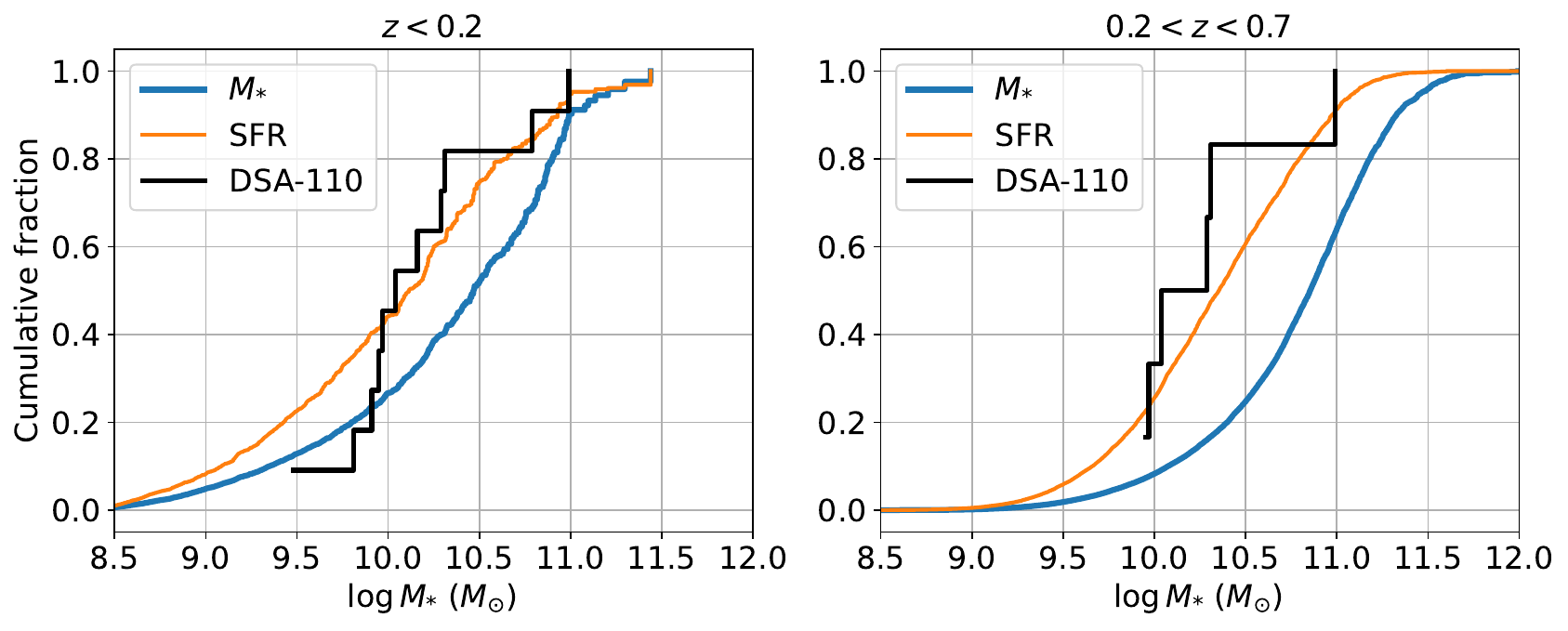}
    \caption{The cumulative fraction of stellar mass ($M_{*}$ -- blue curves), star-formation rate (SFR -- orange curves), and DSA-110 FRBs (black curves) contributed by galaxies above a given stellar mass. Data are shown in two redshift intervals ($z<0.2$ and $0.2<z<0.7$) that evenly split the number of DSA-110 FRBs. The background galaxy sample is from the COSMOS field \citet{2020ApJ...893..111L}, analyzed using identical techniques to the DSA-110 FRB hosts. We have applied the same magnitude limits ($r<23.2$ and $i<23.1$) in the DSA-110 sample to the COSMOS sample.}
    \label{fig:hostmass}
\end{figure}

Figure~\ref{fig:hostmass} shows the cumulative distribution of the hosts in stellar mass, in comparison with the cumulative distributions of stellar mass and star-formation rate of the background galaxy population. The background galaxy population was obtained from the COSMOS sample modeled with \sw{Prospector} \citet{2020ApJ...893..111L} using selections identical to the PS1 magnitude limits for the FRB host sample. Roughly half of the star-formation in the Universe is contributed by galaxies that contribute only 15 -- 30\% of its stellar mass (for $z<0.7$). That distinction, and how it evolves with redshift, allows us to associate FRB formation with specific environments by comparing mass distributions. If FRB occurrence is tied to either current or the accumulation of past star formation, we expect the stellar-mass distribution to trace that of star-formation or stellar mass, respectively. \newtext{Despite the presence of massive, quiescent hosts, the DSA-110 hosts tend to follow the cumulative SFR distribution in both redshift bins. This is consistent with previous FRB host distributions and FRB rate analyses that attribute FRB formation to recent star formation \citep{2021ApJ...907L..31B,2022MNRAS.510L..18J,2023arXiv230205465G}.}\oldtext{shows that FRB hosts with $z<0.2$ show no clear preference for either distribution, but that $0.2<z<0.7$ hosts follow the stellar mass distribution.}

More detailed insight can be gained by associating the occurrence of FRBs to specific stellar-population ages via a delay time distribution, as described below.

\subsection{Delay-Time Distribution} \label{sec:dtd}

FRBs and other transients are produced by sources that are born during periods of star-formation. Galaxies that are actively forming stars are associated with short-lived sources \citep[e.g., core-collapse supernovae;][]{2015A&A...584A..62C}, while early-type galaxies are more likely to be associated with long-lived transient progenitors \citep{2007ApJ...665.1220Z}. The delay-time distribution (DTD) describes the time between the formation of stars and transient events and has been used to study short GRBs \citep{2022ApJ...940L..18Z}, SN Ia \citep{2021MNRAS.506.3330W}, and core-collapse supernovae \citep{2021MNRAS.501.3122C}.

The delay-time distribution is used to statistically infer transient origins. For example, the minimum delay time for core-collapse supernovae was initially assumed to be tied to the lifetime of massive stars ($t_d\approx20-50$\ Myr). However, DTD analysis now supports theoretical modeling that shows binary evolution introduces a tail of $\sim$15\% ``late''  explosions \citep[up to 200 Myr;][]{2012Sci...337..444S,2021MNRAS.501.3122C}.
Compact object binaries are expected to inspiral as they emit gravitational radiation, producing a DTD powerlaw slope $\beta=-1$. For comparison, short GRBs
have a delay time powerlaw slope $\beta=-1.83^{+0.35}_{-0.39}$ \citep[from 200 Myr to 8 Gyr;][]{2022ApJ...940L..18Z} and for SN Ia supernovae
$\beta=-1.13^{+0.04}_{-0.06}$ \citep[assuming $t_{min}=40 \rm{Myr}$;][]{2021MNRAS.506.3330W}.

For the present host galaxy sample, we model the DTD as a powerlaw with a minimum, maximum, and slope for the FRB time relative to the star-formation history of each galaxy \citep[first described in][]{2023arXiv230214782S}. The probability of detecting an FRB is assumed to follow a Poisson distribution, such that for a given host galaxy $i$ and a star formation history posterior sample $j$, the expected rate of FRBs is $\dot{n}_i^j$ at redshift $z_i^j$. 
\newtext{FRB repetition seems to violate the assumption that the delay time is calculated for a single, cataclysmic event. However, FRB source models predict activity on relatively short timescales \citep[$<10^6$\ yr;][]{
2020ApJ...899L..27M,2023MNRAS.525L..22K} or with rapid decay powerlaw \citep[$<-2$;][]{2021ApJ...919...89Y} compared to the typical delay times and slopes. In this case, the chance of discovering an FRB is a reasonable estimate of the maximum of its burst rate (modulated by potential for obscuration by surrounding material). Therefore, we argue that the discovery date can be treated as a once-off event for DTD purposes.}

\begin{figure*}
    \centering
    \includegraphics[width=0.5\textwidth]{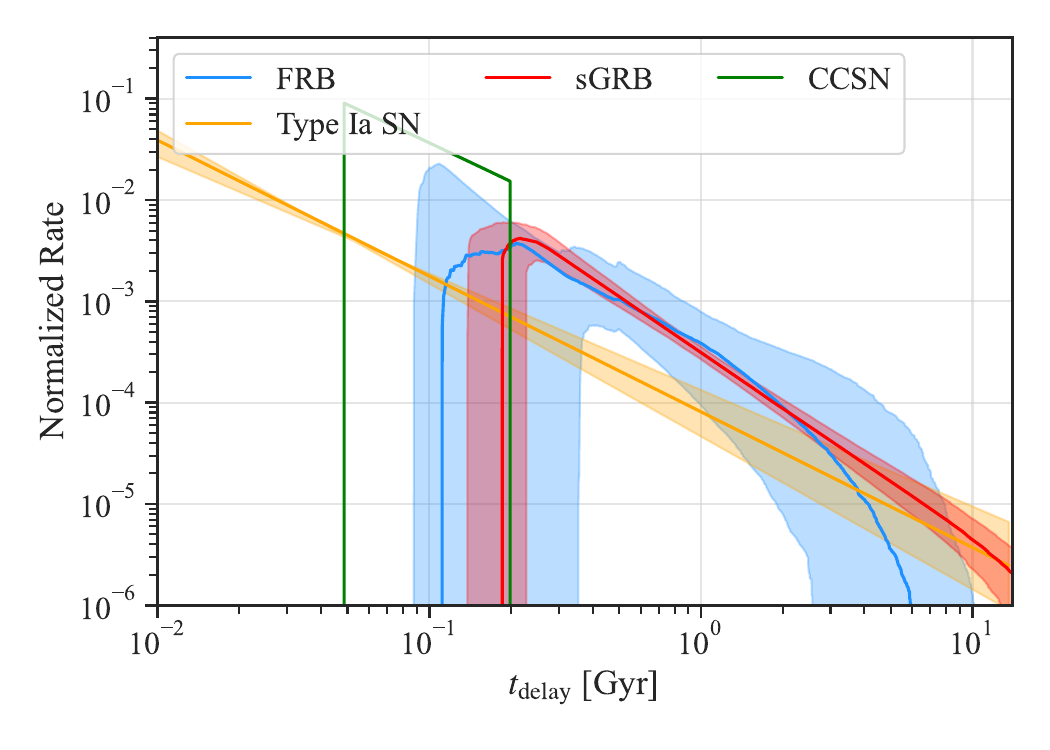}
    \caption{Delay time distributions for galaxies relative to transients discovered in them. The blue line shows our DSA-110 FRB sample, redline shows short GRBs \citep{2022ApJ...940L..18Z}, green line shows core-collapse SNe \citep{2017A&A...601A..29Z}, and yellow line shows SN-Ia \citep{2019ApJ...882...52H}. The FRB DTD is measured non-parametrically from the star-formation history distributions presented here \citep[see also][]{2023arXiv230214782S}.}
    \label{fig:dtd}
\end{figure*}

Figure \ref{fig:dtd} shows the DTD of the current FRB host galaxy sample compared to that of other transient classes. The minimum and maximum delay time is constrained to be 110$^{+250}_{-20}$\ Myr and 6.7$^{+4.56}_{-4.15}$\ Gyr, respectively, and slope $\beta=-1.93^{+1.10}_{-0.71}$. The posterior distribution requires \oldtext{a very wide range of}\newtext{some chance of long} delay times, which supports the conclusions of \citet{2023arXiv230214782S}. While the FRB host sample is small, it looks similar to that of the short gamma-ray burst DTD \newtext{under our assumption of a powerlaw DTD}.

An important caveat to this analysis
\oldtext{First, the traditional DTD analysis assumes that transients are cataclysmic, but we know that some (or all) FRBs can emit multiple bursts and the burst rate may evolve in time \citep[e.g.,][]{2017ApJ...841...14M,2017MNRAS.468.2726K}. }\newtext{is that we} assume a single, stationary Poisson process for the entire FRB population. \newtext{In fact,} there may be multiple kinds of FRB. For these reasons, the slope of the DTD powerlaw should not be directly compared to that of other DTDs. The minimum and maximum of the distribution are likely robust to these caveats. However, we expect the estimate of $t_{\mathrm{min}}$ to be affected by the lack of data constraining recent star formation history for some hosts.

Despite these caveats, the DTD analysis provides useful context to the binary framing of whether FRB hosts trace star formation or stellar mass. We find FRBs can occur with either a short \newtext{($\sim100$\ Myr)} or long \newtext{($\gtrsim2$\ Gyr)} delay from star formation. \newtext{This could be interpreted as a single FRB population with a powerlaw spanning a range of delay times or multiple populations and an overall rate dominated by short delay times.}\oldtext{, but the presence of long delays produces a stronger correlation with the integrated history of star formation of a galaxy (i.e., stellar mass).} \newtext{Some} modeling of the larger sample of FRB DMs and redshifts show that the rate evolves with redshift similarly to the mean star formation rate of the universe, \oldtext{albeit with large errors}\newtext{, which is consistent with the association of FRBs to mostly star-forming hosts in the present and previous samples} \citep{2019ApJ...883...40L,2021ApJ...907L..31B,2022MNRAS.510L..18J}.
\newtext{Alternatively, analysis of the DM-implied distance distribution of CHIME/FRB discoveries suggests formation via a delayed channel, which is consistent with the wide DTD measured here \citep{2020MNRAS.498.3927H,2022ApJ...924L..14Z}.}
It should eventually be possible to \oldtext{see}\newtext{resolve this tension through} deviations of the evolution of the FRB rate from that of star-formation, which scales as $(1+z)^{2.7}$ in the local universe. Specifically, long delays shift the rate peak to later times and flatten its evolution. This is evident in other transient populations with wide delay time distributions, such as SN Ia and short GRBs \citep{2008MNRAS.388..829G,2022ApJ...940L..18Z}.

\section{Conclusions}
\label{sec:con}

We presented a sample of 11 FRBs discovered by the DSA-110 and selected on their association to host galaxies. This represents the largest and most uniform sample of FRB host galaxies. The FRBs are not known to repeat and are fainter and narrower in time than previous large FRB samples. Despite that, the bursts are phenomenologically similar to previous FRB discoveries from wide-field survey telescopes, such as CHIME and ASKAP. The new host galaxy sample supports prior work in associating FRBs to galaxies with active star formation. However, we find a wide delay-time distribution for FRBs relative to star-formation. \oldtext{In the DSA-110 sample, we see this as a FRB host stellar mass distribution that matches the field galaxy stellar mass distribution.}

The wide delay time distribution requires that \oldtext{either (1) the}\newtext{at least some sources of} FRBs are formed over a wide range of times relative to star formation.
\oldtext{or (2) that they are formed during star formation and can emit up to a few Gyr after formation. The latter scenario is disfavored if bursts are powered by spin or magnetic fields in neutron stars \citep{2017ApJ...841...14M,2022MNRAS.510.1867L}. Under the former scenario, single or multiple formation channels are allowed.}
Binary or dynamical formation channels \citep[e.g.,][]{2019ApJ...886..110M,2021ApJ...917L..11K} are particularly attractive, because they can form over a range of delays and naturally explain some burst phenomenology, such as periodic activity cycles. It remains difficult to distinguish between the single and multiple FRB formation channels because burst properties only weakly correlate with environmental properties. With this new sample, we strengthen the evidence for repeating FRBs to be associated with PRS, which, if true, suggests that activity is causally related to the PRS emission.

The DSA-110 continues to detect FRBs and associate them to host galaxies. Future DSA-110 publications will use large, uniform samples of burst spectra and polarization to classify sources and characterize their environments. Larger host galaxy samples will include morphological and offset analysis that can test formation channels. It may eventually be possible to detect enough hosts to model the FRB rate with multiple formation channels, as has been done for SN Ia \citep{2006ApJ...648..868S}. Furthermore, FRB localization samples are still small, such that any effects or subclasses that occur in less than 1\% of the population remain to be discovered.

%

\begin{acknowledgements}

We thank the OVRO staff for making this science possible through epidemics, fires, floods, and other disasters. The observatory is located on the ancestral homelands of the Big Pine Paiute Tribe of the Owens Valley. We recognize and acknowledge the historical and cultural significance of these lands to members of the Tribe.

The DSA-110 is supported by the National Science Foundation Mid-Scale Innovations Program in Astronomical Sciences (MSIP) under grant AST-1836018. Some of the data presented herein were obtained at the W. M. Keck Observatory, which is operated as a scientific partnership among the California Institute of Technology, the University of California and the National Aeronautics and Space Administration. The Observatory was made possible by the generous financial support of the W. M. Keck Foundation. 

Some of the data presented herein were obtained at the W. M. Keck Observatory, which is operated as a scientific partnership among the California Institute of Technology, the University of California and the National Aeronautics and Space Administration. The Observatory was made possible by the generous financial support of the W. M. Keck Foundation. The authors wish to recognize and acknowledge the very significant cultural role and reverence that the summit of Maunakea has always had within the indigenous Hawaiian community. We are most fortunate to have the opportunity to conduct observations from this mountain. 

We acknowledge use of the VLA calibrator manual and the radio fundamental catalog. This research has made use of NASA’s Astrophysics Data System. 

The Pan-STARRS1 Surveys (PS1) and the PS1 public science archive have been made possible through contributions by the Institute for Astronomy, the University of Hawaii, the Pan-STARRS Project Office, the Max-Planck Society and its participating institutes, the Max Planck Institute for Astronomy, Heidelberg and the Max Planck Institute for Extraterrestrial Physics, Garching, The Johns Hopkins University, Durham University, the University of Edinburgh, the Queen's University Belfast, the Harvard-Smithsonian Center for Astrophysics, the Las Cumbres Observatory Global Telescope Network Incorporated, the National Central University of Taiwan, the Space Telescope Science Institute, the National Aeronautics and Space Administration under Grant No. NNX08AR22G issued through the Planetary Science Division of the NASA Science Mission Directorate, the National Science Foundation Grant No. AST-1238877, the University of Maryland, Eotvos Lorand University (ELTE), the Los Alamos National Laboratory, and the Gordon and Betty Moore Foundation.

The Legacy Surveys consist of three individual and complementary projects: the Dark Energy Camera Legacy Survey (DECaLS; Proposal ID \#2014B-0404; PIs: David Schlegel and Arjun Dey), the Beijing-Arizona Sky Survey (BASS; NOAO Prop. ID \#2015A-0801; PIs: Zhou Xu and Xiaohui Fan), and the Mayall z-band Legacy Survey (MzLS; Prop. ID \#2016A-0453; PI: Arjun Dey). DECaLS, BASS and MzLS together include data obtained, respectively, at the Blanco telescope, Cerro Tololo Inter-American Observatory, NSF’s NOIRLab; the Bok telescope, Steward Observatory, University of Arizona; and the Mayall telescope, Kitt Peak National Observatory, NOIRLab. Pipeline processing and analyses of the data were supported by NOIRLab and the Lawrence Berkeley National Laboratory (LBNL). The Legacy Surveys project is honored to be permitted to conduct astronomical research on Iolkam Du’ag (Kitt Peak), a mountain with particular significance to the Tohono O’odham Nation.

\newtext{This research uses services or data provided by the Astro Data Lab at NSF's National Optical-Infrared Astronomy Research Laboratory.} NOIRLab is operated by the Association of Universities for Research in Astronomy (AURA) under a cooperative agreement with the National Science Foundation. LBNL is managed by the Regents of the University of California under contract to the U.S. Department of Energy.

The Legacy Surveys imaging of the DESI footprint is supported by the Director, Office of Science, Office of High Energy Physics of the U.S. Department of Energy under Contract No. DE-AC02-05CH1123, by the National Energy Research Scientific Computing Center, a DOE Office of Science User Facility under the same contract; and by the U.S. National Science Foundation, Division of Astronomical Sciences under Contract No. AST-0950945 to NOAO.
\end{acknowledgements}

\software{\sw{astropy} \citep{2013A&A...558A..33A,2018AJ....156..123A,2022ApJ...935..167A}, \sw{astroquery} \citep{2019AJ....157...98G}, \oldtext{\sw{astro-datalab} \citep{}}, astropath \citep{2021ApJ...911...95A}, \sw{emcee} \citep{2013PASP..125..306F}, \sw{lpipe} \citep{2019PASP..131h4503P}, \sw{Prospector}~\citep{2021ApJS..254...22J}, CASA \citep{2022PASP..134k4501C}, Heimdall \citep{2012PhDT.......465B}, wsclean \citep{2014MNRAS.444..606O}, \sw{dynesty \newtext{(v2.1.0)}}  \citep{2020MNRAS.493.3132S}, \sw{Bilby} \citep{2019ApJS..241...27A,2020MNRAS.499.3295R} , \newtext{FRB \citep{10.5281/zenodo.8125230}, \sw{photutils} \citep{10.5281/zenodo.1035865}}}

\facility{Hale, VLA, Keck:I (LRIS), Keck:II (ESI), DSA-110, PS1}

\begin{appendix}

\section{Spectroscopy and Spectral Modeling}
\label{sec:spres}

Table \ref{tab:details} summarizes the photometric and spectroscopic measurements made for the FRB host galaxies. Figure \ref{fig:SED_fits} shows the modeling of the host galaxy photometry and spectroscopy.

\begin{table}
\begin{tabular}{l|cc}
FRB name & Photometry & Spectroscopy \\ \hline
\frbzach & PS1 grizy, 2MASS HKs, ALLWISE w1w2 & Keck/LRIS \\
\frbalex & PS1 grizy, WIRC JH & Hale/DBSP \\
\frbwhitney & PS1 riz & Keck/LRIS \\
\frbmark & PS1 grizy, 2MASS HKs, ALLWISE w1w2 & Hale/DBSP \\
\frbquincy & PS1 grizy, WIRC JH & Keck/LRIS \\
\frboran & PS1 rizy, WIRC JH, ALLWISE w1w2 & Keck/LRIS \\
\frbjackie & PS1 grizy, 2MASS HKs, ALLWISE w1w2 & Keck/LRIS\tablenotemark{a} \\
\frbansel & PS1 grizy & Keck/LRIS\tablenotemark{a} \\
\frbelektra & PS1 grizy & Keck/LRIS\tablenotemark{a} \\
\frbetienne & PS1 grizy & Hale/DBSP \\
\frbjuan & PS1 grizy, 2MASS HKs, ALLWISE w1w2 & Keck/LRIS\tablenotemark{a} \\
\end{tabular}
\caption{Photometric and spectroscopic observations of the DSA-110 FRB host galaxy sample}
\tablenotetext{a}{Observed with blue side.}
\label{tab:details}
\end{table}

\begin{table}
\begin{tabular}{l|cccc}
FRB name & H$\alpha$  & H$\beta$ & [OIII] 5007 & [NII] 6584 \\ \hline
\frbzach & 8.65$\times10^{-16}$ & 5.11$\times10^{-17}$ & 5(1)$\times10^{-17}$ & 1.81(7)$\times10^{-16}$ \\
\frbalex & 5.9(1)$\times10^{-15}$ & & & 2.3(1)$\times10^{-15}$ \\
\frbwhitney & 8.924$\times10^{-17}$ & 1.93$\times10^{-17}$ & 2.6(4)$\times10^{-17}$ & 2.3(6)$\times10^{-17}$ \\
\frbmark\tablenotemark{a} & 2.00(3)$\times10^{-14}$ & & & 1.46(4)$\times10^{-14}$ \\
\frbquincy & & 4.89$\times10^{-17}$ & 2.1(5)$\times10^{-17}$ & \\
\frboran & 1.18$\times10{-16}$ & 4.98$\times10^{-18}$ & & 4.9(5)$\times10^{-17}$ \\
\frbjackie\tablenotemark{b} & & & & \\
\frbansel & & 1.23(3)$\times10^{-16}$ & 4.0(2)$\times10^{-17}$ & \\
\frbelektra & & 9.0(2)$\times10^{-17}$ & 7.5(2)$\times10^{-17}$ & \\
\frbetienne & 1.08$\times10^{-15}$ & 3.41$\times10^{-16}$ & 1.5(3)$\times10^{-16}$ & 3.1(1)$\times10^{-16}$ \\
\frbjuan & & 1.00(4)$\times10^{-16}$ & 2.0(3)$\times10^{-17}$ &  \\
\end{tabular}
\caption{Host galaxy line fluxes for four transitions used in BPT diagram (Figure \ref{fig:hostprops}). \newtext{Line fluxes are not corrected for Galactic or host extinction.}}
\tablenotetext{a}{Lines measured in nucleus of host \citet[see also ][]{2023ApJ...949L...3R}.}
\tablenotetext{b}{No emission lines detected, but redshift measured from absorption lines.}
\label{tab:lines}
\end{table}

\begin{figure*}
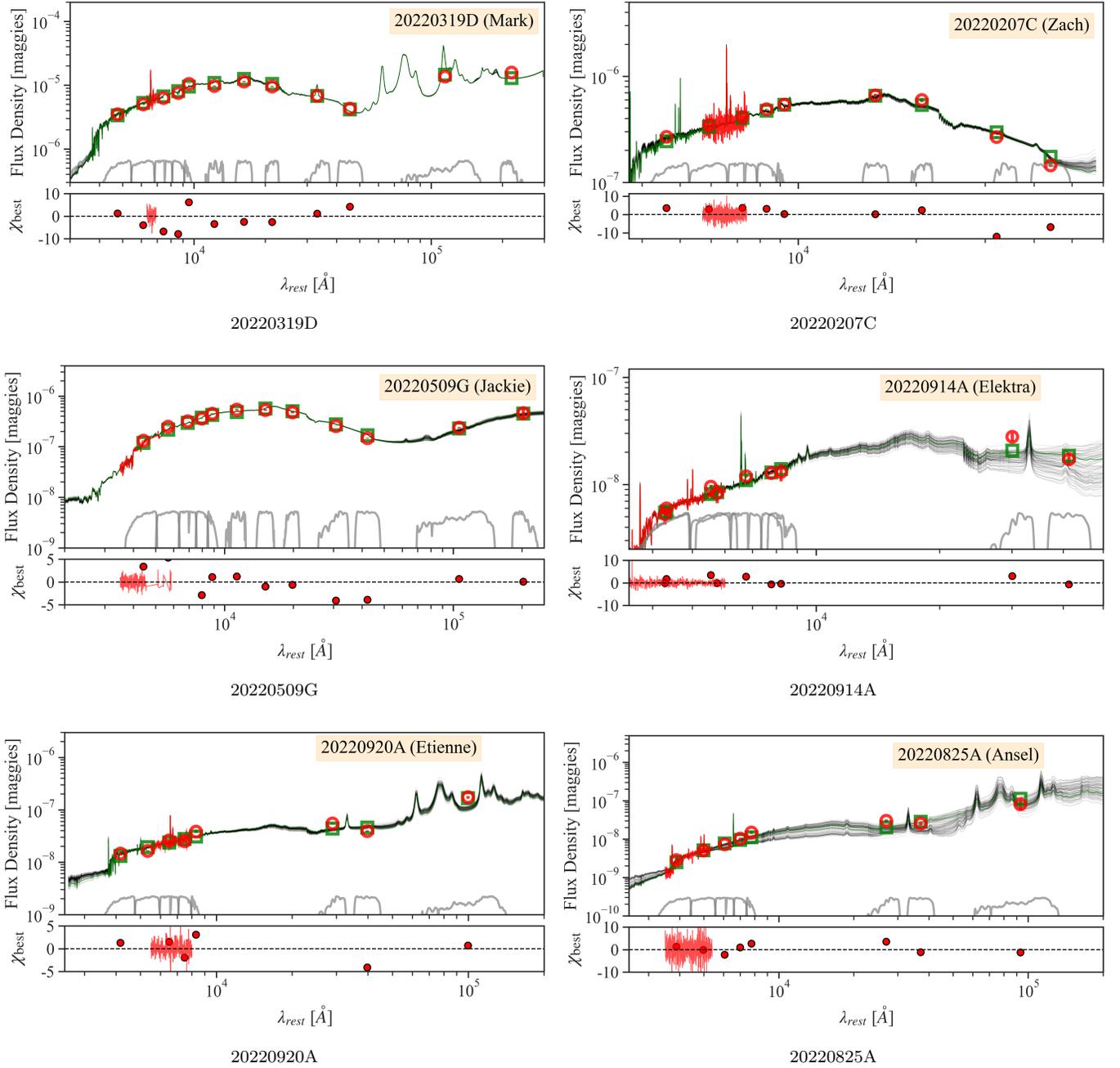

    \gridline{\fig{hostmodeling/mark_sed_fit.png}{0.5\linewidth}{\frbmark}
              \fig{hostmodeling/zach_sed_fit.png}{0.5\linewidth}{\frbzach}}
    \gridline{\fig{hostmodeling/jackie_sed_fit.png}{0.5\linewidth}{\frbjackie}
              \fig{hostmodeling/elektra_sed_fit.png}{0.5\linewidth}{\frbelektra}}
    \gridline{\fig{hostmodeling/etienne_sed_fit.png}{0.5\linewidth}{\frbetienne}
              \fig{hostmodeling/ansel_sed_fit.png}{0.5\linewidth}{\frbansel}}
    \caption{Optical and infrared spectral energy distribution (SED) for FRB host galaxies (Part 1 of 2). Each figure has two panels with measurements, model, and bandpasses in the top and residuals in the bottom. The top panel show measurements in red, the best model values at the wavelength/band of the measurement in green, and bandpasses in grey. Black lines show a set of random models drawn from the best posterior model distribution. The top right of each figure shows the reduced chi-squared of the best model fit.}
    \label{fig:SED_fits}
\end{figure*}

\begin{figure*}
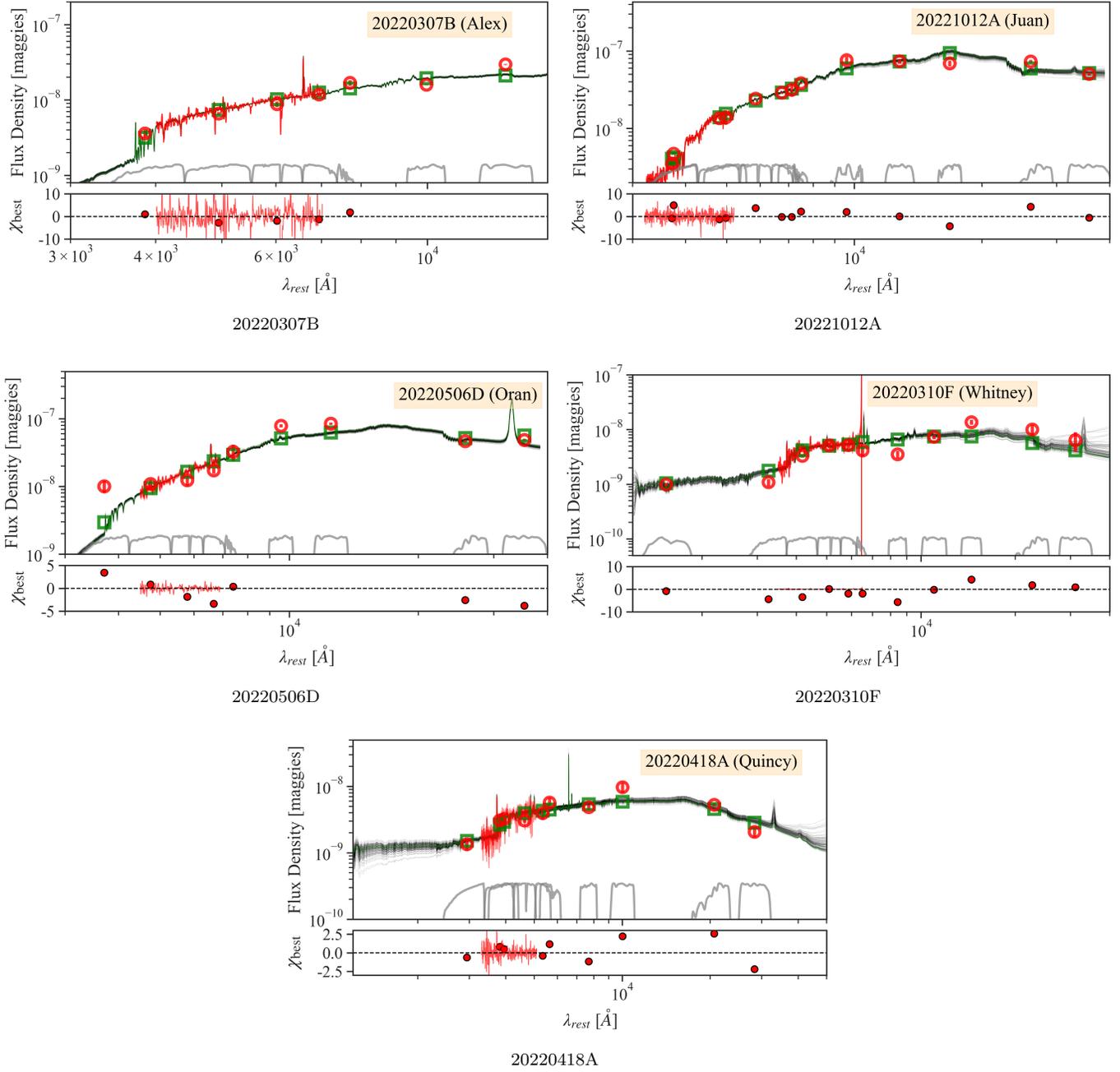

    \gridline{\fig{hostmodeling/alex_sed_fit.png}{0.5\linewidth}{\frbalex}
              \fig{hostmodeling/juan_sed_fit.png}{0.5\linewidth}{\frbjuan}}
    \gridline{\fig{hostmodeling/oran_sed_fit.png}{0.5\linewidth}{\frboran}
              \fig{hostmodeling/whitney_sed_fit.png}{0.5\linewidth}{\frbwhitney}}
    \gridline{\fig{hostmodeling/quincy_sed_fit.png}{0.5\linewidth}{\frbquincy}}
    \caption{Optical and infrared spectral energy distribution (SED) for FRB host galaxies (Part 2 of 2).}
    \label{fig:SED_fits}
\end{figure*}

\begin{figure*}
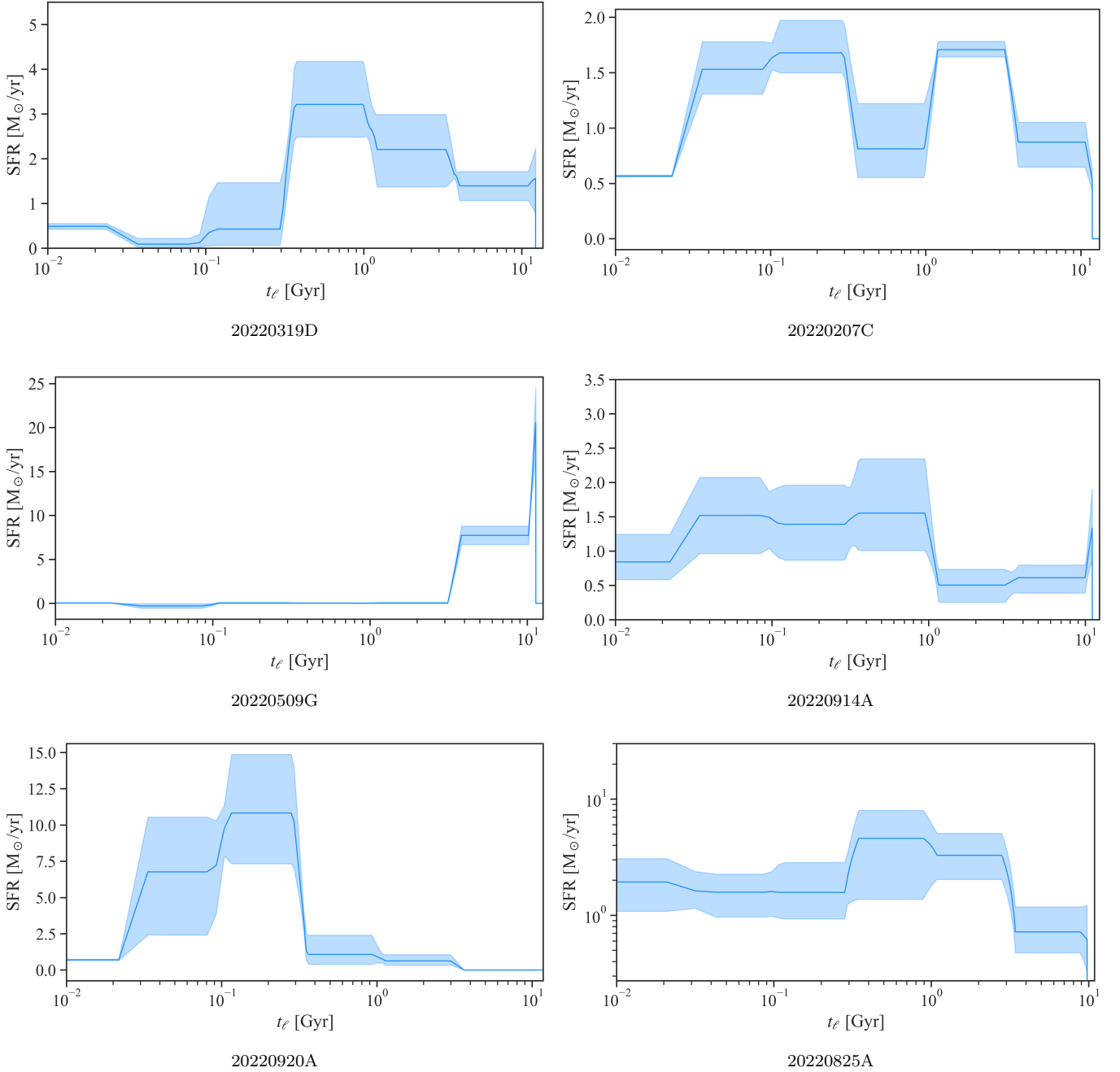

    \gridline{\fig{hostmodeling/mark_sfh.png}{0.5\linewidth}{\frbmark}
              \fig{hostmodeling/zach_sfh.png}{0.5\linewidth}{\frbzach}}
    \gridline{\fig{hostmodeling/jackie_sfh.png}{0.5\linewidth}{\frbjackie}
              \fig{hostmodeling/elektra_sfh.png}{0.5\linewidth}{\frbelektra}}
    \gridline{\fig{hostmodeling/etienne_sfh.png}{0.5\linewidth}{\frbetienne}
              \fig{hostmodeling/ansel_sfh.png}{0.5\linewidth}{\frbansel}}
    \caption{Star formation history for FRB host galaxies (Part 1 of 2). Prospector modeling (\S \ref{sec:sed}) uses seven, logarithmically-spaced bins with a continuity prior that favors smoother changes in star-formation rate between bins.}
    \label{fig:SFH_fits}
\end{figure*}

\begin{figure*}
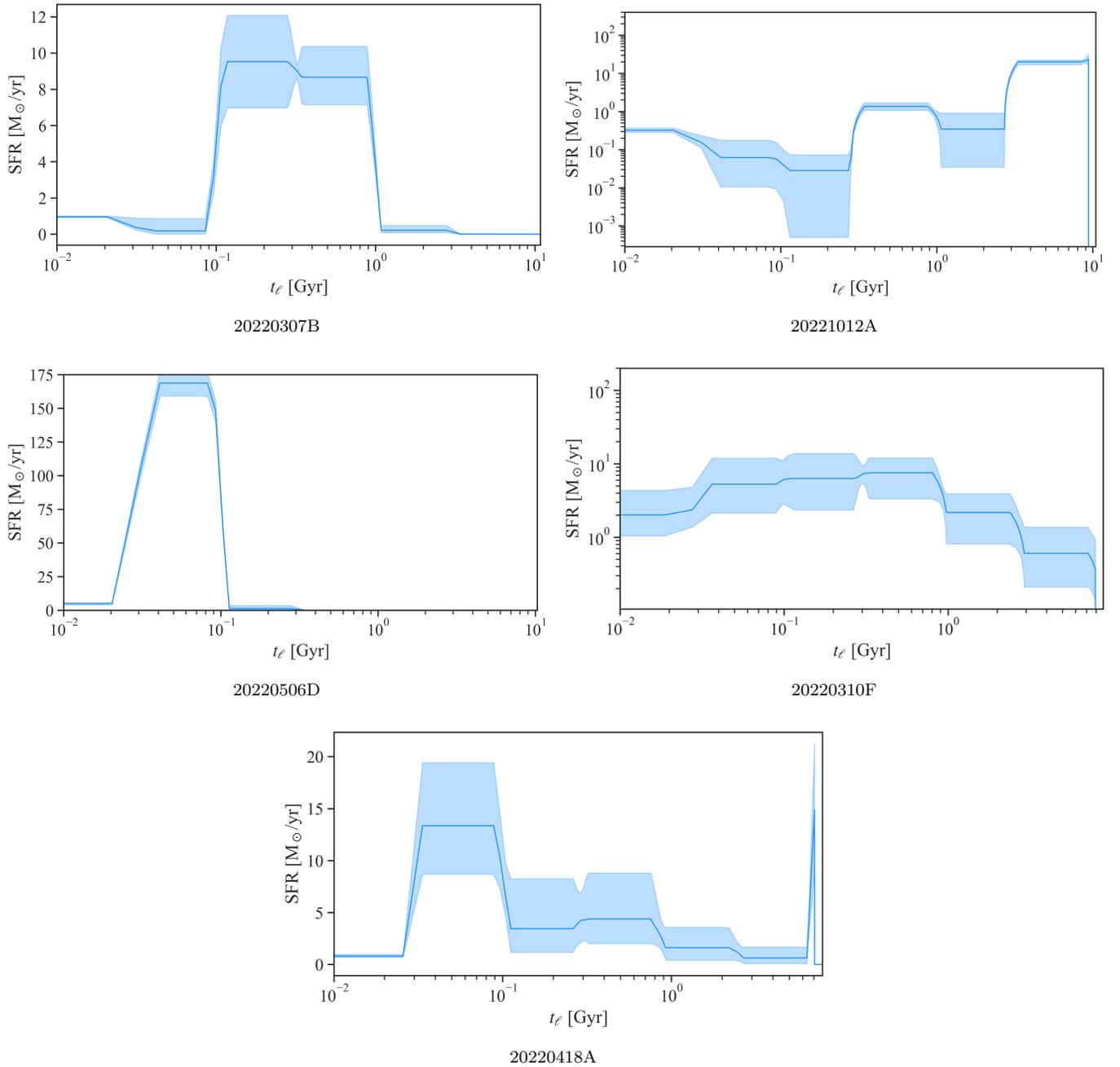

    \gridline{\fig{hostmodeling/alex_sfh.png}{0.5\linewidth}{\frbalex}
              \fig{hostmodeling/juan_sfh.png}{0.5\linewidth}{\frbjuan}}
    \gridline{\fig{hostmodeling/oran_sfh.png}{0.5\linewidth}{\frboran}
              \fig{hostmodeling/whitney_sfh.png}{0.5\linewidth}{\frbwhitney}}
    \gridline{\fig{hostmodeling/quincy_sfh.png}{0.5\linewidth}{\frbquincy}}
    \caption{Star formation history for FRB host galaxies (Part 2 of 2).}
    \label{fig:SFH_fits}
\end{figure*}

\end{appendix}

\bibliography{fasttransients}{}
\bibliographystyle{aasjournal}

\end{document}